\documentclass[preprint,preprintnumbers,amsmath,amssymb]{revtex4}
\usepackage[abs]{overpic}
\usepackage{graphicx,subfigure}
\usepackage{amsmath}
\usepackage{slashed}
\begin{document}
\preprint{HUPD1102}
\def\tbr{\textcolor{red}}
\def\tcr{\textcolor{red}}
\def\ov{\overline}
\def\nn{\nonumber}
\def\f{\frac}
\def\beq{\begin{equation}}
\def\eeq{\end{equation}}
\def\bea{\begin{eqnarray}}
\def\eea{\end{eqnarray}}
\def\bsub{\begin{subequations}}
\def\esub{\end{subequations}}
\def\dc{\stackrel{\leftrightarrow}{\partial}}
\def\ynu{y_{\nu}}
\def\ydu{y_{\triangle}}
\def\ynut{{y_{\nu}}^T}
\def\ynuv{y_{\nu}\frac{v}{\sqrt{2}}}
\def\ynuvt{{\ynut}\frac{v}{\sqrt{2}}}
\def\d{\partial}
\title{The form factors of
$\tau \to K \pi (\eta) \nu$ and 
the predictions for CP violation beyond the standard model} 
\author{Daiji Kimura$\, ^{\rm a}$, Kang Young Lee$\, ^{\rm b}$ and 
Takuya Morozumi$\, ^{\rm c}$}
\affiliation{$^{\rm a}\, $
Faculty of Engineering, Kinki University,
Takaya, Higashi-Hiroshima, 739-2116, Japan, \\
%
$^{\rm b}$\, Department of Physics Education and Education Research Institute,
Gyeongsang National University,
Jinju 660-701, Korea,
$^{\rm c}$\, Graduate School of Science, Hiroshima University,
Higashi-Hiroshima, 739-8526, Japan
}
\def\nn{\nonumber}
\def\beq{\begin{equation}}
\def\eeq{\end{equation}}
\def\bei{\begin{itemize}}
\def\eei{\end{itemize}}
\def\bea{\begin{eqnarray}}
\def\eea{\end{eqnarray}}
\def\ynu{y_{\nu}}
\def\ydu{y_{\triangle}}
\def\ynut{{y_{\nu}}^T}
\def\ynuv{y_{\nu}\frac{v}{\sqrt{2}}}
\def\ynuvt{{\ynut}\frac{v}{\sqrt{2}}}
\def\s{\partial \hspace{-.47em}/}
\def\ad{\overleftrightarrow{\partial}}
\def\ss{s \hspace{-.47em}/}
\def\pp{p \hspace{-.47em}/}
\def\bos{\boldsymbol}
\begin{abstract}
We study the hadronic form factors of $\tau$ lepton
decays $\tau \to K \pi(\eta) \nu$.
We compute one loop corrections
to the form factors
using the chiral Lagrangian including vector mesons.
The counterterms which subtract the divergence of the one-loop amplitudes
 are determined
by using background field method.
In the vector form factor, $K^\ast$ resonance behavior is reproduced
because the diagram with a vector meson propagator is included. 
We fit the data of the hadronic invariant mass spectrum
measured by Belle by determining some of the counterterms of the Lagrangian.
Besides the hadronic invariant mass spectrum,
the forward-backward asymmetry is predicted.
We also study the effect of CP violation of a two Higgs doublet model.
In the model, CP violation of the neutral Higgs sector generates the
mixing of CP even Higgs and CP odd Higgs. We show how the
mixing leads to the direct CP violation of the
$\tau$ decays and predict the CP violation of the forward-backward asymmetry.

\end{abstract}
\maketitle
\section{Introduction}
The tau hadronic decays are unique as the decays can be useful
for the search for the new CP violation beyond the standard model
\cite{Nelson:1993zv,Kilian:1994ub, Tsai:1994rc, Choi:1994ch, 
Finkemeier:1996dh, Kuhn:1996dv, Choi:1998yx, Datta:2006kd, Kiers:2008mv,
Kimura:2008gh,Kimura:2010pa, LopezCastro:2009da} . 
CP violation in tau lepton semileptonic decays has been searched
in $\tau \to \pi \pi \nu $ \cite{Avery:2001md} and $\tau \to K \pi \nu$ \cite{Bonvicini:2001xz, Bischofberger:2011pw} modes.
Recently, Belle \cite{Bischofberger:2011pw} puts constraint
on the CP violation parameter of three Higgs doublet model
with their latest data. Babar \cite{BABAR:2011aa} also searched CP violation 
of $\tau \to K_s \pi^- \nu (n) \pi^0$ $(n>0)$ and obtains non-zero CP asymmetry
which sign is opposite to the standard model prediction \cite{Bigi:2005ts},\cite{
Grossman:2011zk}.
In $\tau$ lepton system, another CP
violating observable, e.d.m.
(electric dipole moment) is 
also searched \cite{Inami:2002ah}.

To predict the direct CP violation of the hadronic
$\tau$ decays, the strong phase shifts 
are important quantities and the quantitative prediction
on the strong phase shifts is necessary when extracting
the weak CP violating phases from the experimentally 
observed CP asymmetries \cite{Kuhn:1996dv, Kimura:2008gh, Kimura:2010pa}. 
This is a reason why we study
the hadronic form factors.

  The hadronic form factors for the decay $\tau \to K \pi \nu$
and SU(2) isospin vector form factors
have been studied with various methods. The common future of them
is the effects of vector mesons ($\rho, K^*$) and higher resonances are included.
In Ref.\cite{Finkemeier:1996dh, Fajfer:1999hh}, 
vector dominance models
are studied.
One loop corrections to the SU(2) vector form factor are studied with
the resonance chiral theory in \cite{Rosell:2004mn}.
In Ref.
\cite{Jamin:2006tk,Moussallam:2007qc,Boito:2008fq,Bernard:2009zm}
the $\tau \to K \pi \nu $ form factors are predicted using the chiral theory combined  
with the dispersion relations. 
In our previous study, we use the resonance chiral Lagrangian including the one loop corrections to the self-energy of resonance \cite{Kimura:2008gh}.
We also note in the experimental study \cite{Belle:2007rf}, the Breit-Wigner
form for several resonances is used to fit the data of the hadronic
invariant mass spectrum. 

In this paper, we use different approach from the previous
study \cite{Kimura:2008gh}. By using the chiral Lagrangian including 
vector resonance \cite{Bando:1984ej}, we compute the one loop corrections
of pseudoscalar mesons to the form factors. Including the Feynman
diagrams with a
vector meson propagator, one can 
reproduce the resonance behavior near the poles
of the resonances while keeping 
prediction at threshold region consistent with the chiral symmetry.
Our Lagrangian includes several new counterterms related to vector mesons
in addition to the counterterms which are present in chiral perturbation
theory. The coefficients of the counterterms are different from the
chiral perturbation theory.

Since Belle and Babar reported the precise measurements of
the branching fractions of $ \tau^- \to K_s \pi^- \nu $ 
\cite{Belle:2007rf}, $\tau^- \to K^- \pi^0 \nu$ \cite{Aubert:2007jh}
and $\tau^- \to K^- \eta \nu$ \cite{Inami:2008ar, delAmoSanchez:2010pc}, 
we can compare our prediction of the hadronic
invariant mass distribution with the experimental data.
We have determined the finite parts of the coefficients  of counterterms
so that the hadronic invariant mass spectrum is reproduced.

Once the form factors are fixed, one can use them for predictions
of various distributions within the standard model (SM) and beyond.
We first compute the angular distribution and the forward-backward 
asymmetry (FBA) for $\tau \to K \pi \nu$ and
$\tau \to K \eta \nu$ in SM \cite{Beldjoudi:1994hi}.
Furthermore, CP violation of FBA is predicted with a 
two Higgs doublet model.  In type II two Higgs doublet model, within the
tree level approximation,
the charged Higgs couplings with quarks and leptons are written in terms
of Cabibbo-Kobayashi-Maskawa (CKM) matrix. However, if we take into account
the one loop corrections to the masses of quarks 
and leptons due to the neutral Higgs exchanged diagrams, CP violation of the neutral Higgs sector becomes a new source 
of CP violation of the charged Higgs Yukawa couplings.  We show the CP 
violation of the type II two Higgs doublet model can be 
probed by the direct CP violation of the $\tau$ hadronic decays.

The paper is organized as follows. 
In section II, we show the hadronic chiral Lagrangian
with the vector resonances 
which are relevant for the form factors. 
The counterterms are also given.
In section III, we compute the form factors.
In section IV, we calculate the hadronic invariant 
mass spectra of the decays $\tau \to K \pi(\eta) \nu$. 
The spectra are compared with the experimental
data and the FBAs are predicted.
In section V, we explain how CP violation in neutral Higgs
sector reveals itself in the charged Higgs Yukawa couplings
in a two Higgs doublet model.
We also calculate the CP violation of FBA of
the hadronic $\tau$ decays and the numerical result is presented.
Section VI is devoted to discussion and summary.
In appendix, we give some details of the derivation of the
formulae used in the text.
\section{Chiral Lagrangian with vector mesons}
The leading order $O(p^2)$ of chiral Lagrangian with $\eta_0-\eta_8$
mixing term and vector meson mass term is given by,
\bea
{\cal L}&=&\frac{f^2}{4} {\rm Tr} (D_L U D_L U^{\dagger})
+B {\rm Tr}[M (U+U^{\dagger})] -ig_{2p}{\rm Tr}(\xi M\xi 
-\xi^\dagger M\xi^\dagger)\cdot \eta_0 \nn \\
&+&\frac{1}{2} 
\partial_\mu \eta_0 \partial^\mu \eta_0-\frac{M_0^2}{2}\eta_0^2
+ M_{V}^2 {\rm Tr}\left[(V_{\mu}-\frac{\alpha_{\mu}}{g})^2 \right] ,  
\label{eq:lag} 
\eea
where $U$ is the chiral field which is given as $U=\exp(2i\pi/f)=\xi^2$.
$\pi$ is SU(3) octet pseudoscalar and $B$ is a constant parameter.
$\eta_0$ is U(1)$_A$ pseudoscalar
of which mass is denoted by $M_0$ and $g_{2p}$ is the coupling
for $\eta_0-\eta_8$ mixing.  
The covariant derivative for the chiral field $U$ is given by,
\begin{equation}
D_{L\mu} U = (\partial_\mu + iA_{L\mu})U , 
\end{equation}
where the external gauge field of SU(3)$_L$ denoted by $A_L$ is introduced.
$V_\mu$ is the vector nonets and $\alpha_\mu$ is defined as,
\bea
\alpha_{\mu}&=&\frac{\xi^\dagger D_{L \mu} \xi+\xi \partial_\mu \xi^\dagger}
{2i}, \nn \\
&=& \alpha^0_{\mu}+ \frac{\xi^\dagger A_{L \mu} \xi}{2}.
\eea
The form of the mass term of  vector mesons is identical to
the that of the unitary gauge fixed version of 
hidden local symmetry approach \cite{Bando:1984ej,Tanabashi:1993np,Harada:2003jx}.
The kinetic term
of the vector mesons is not included in the leading order.
This treatment is important when including loop corrections
in a systematic way.
Note that we have added the chiral
breaking term by $M={\rm diag}(m_u,m_d,m_s)$
for the pseudoscalar mesons. 
The chiral breaking term $\chi$
in the isospin limit can be written in terms of  
the masses of $\pi$ and $K$ mesons as,
\bea
\chi&=&\frac{4 B M}{f^2}, \nn \\
     &=&\left(\begin{array}{ccc}
m_\pi^2 & 0 & 0 \\
0  & m_{\pi}^2 & 0 \\
0  & 0 & 2 m_K^2-m_\pi^2 
\end{array} \right).
\eea 
\subsection{$\eta-\eta^\prime$ mixing}
When computing the form factors for $\tau \to K \eta^{(\prime)} \nu$,
they are sensitive to the mixing angle of $\eta$ and $\eta^\prime$,
We first summarize the mixing of the octet and singlet pseudoscalar 
meson at one-loop order. 
The self-energy correction for $\eta_0$ and $\eta_8$ sector in one 
loop is computed with the interaction terms shown in Appendix D,
\bea
\frac{1}{2}
\begin{pmatrix}\eta_8 & \eta_0 \end{pmatrix}
\begin{pmatrix}
z^{-1}_{88} p^2-M_{88}^2-\delta M^2_{88} & M_{08}^2 +\delta M^2_{08}\\
M_{08}^2+ \delta M^2_{08} & p^2-M^2_{0}
\end{pmatrix}
\begin{pmatrix}\eta_8 \\ \eta_0 \end{pmatrix},
\label{eq:self}
\eea
where the tree level mass squared matrix 
elements $M_{88}^2$ and $M_{08}^2$
are 
given by $\frac{4 m_K^2-m_\pi^2}{3}$ and 
$\frac{2fg_{2p}}{B \sqrt{3}} (m_K^2-m_\pi^2)$
respectively. 
$z_{88}-1$, $\delta M^2_{08}$, and $\delta M^2_{88}$
are one loop corrections and they are given by,
\bea
z_{88}&=& 1+6 c \mu_K-\frac{8L_4^r}{f^2}(2 m_K^2+m_\pi^2)-\frac{8L_5^r}{f^2}\frac{4 m_K^2-m_\pi^2}{3}, \nn \\
\delta M^2_{08}&=& -\frac{g_{2p}f}{\sqrt{3}B}
\Bigl{[} \frac{8m_K^2-5m_\pi^2}{3} \mu_{\eta_8}+2(4m_K^2-3m_\pi^2) \mu_K
+6 \mu_\pi m_\pi^2 \nn \\
&+&4 (2m_K^2+m_\pi^2)  t_3^r + 8 m_K^2 t_5^r \Bigr{]}, 
\nn \\
\delta M^2_{88}&=& \frac{16m_K^2-7m_\pi^2}{9} \mu_{\eta_8}
+\frac{16 m_K^2-6m_\pi^2}{3} \mu_K +m_\pi^2 \mu_\pi \nn \\
&+& \frac{2 g_{2p} f}{B} \cos \bar{\theta}_{08} \sin \bar{\theta}_{08} 
(\mu_{\eta^\prime}-\mu_{\eta}) \frac{5 m_\pi^2-8m_K^2}{3 \sqrt{3}}\nn \\
&+&16 L_6^r \frac{(2m_K^2+m_\pi^2)(4m_K^2-m_\pi^2)}{3f^2}
+32 L_8^r \frac{m_\pi^4+2(2m_K^2-m_\pi^2)^2}{6f^2},
\label{eq:z88}
\eea
where $c=1-\frac{M_V^2}{g^2f^2}$ and $\mu_P$ denotes $\frac{m_P^2}{32 \pi^2 f^2} \log \frac{m_P^2}{\mu^2}, (P=\pi,K,\eta,\eta^\prime)$. $\mu$ denotes
 renormalization scale. We also introduce the notation; 
$\mu_{\eta_8}=\mu_{\eta}\cos^2\theta_{08} + \mu_{\eta'}\sin^2\theta_{08}.$
$t^r_i (i=3,5) $ and $L^r_i(i=4,5,6,8)$ are the finite counterterms which are defined in Eq.(\ref{eq:fullcounter}) and Eq.(\ref{eq:finitecounter}).
$\bar{\theta}_{08}$ denotes the mixing angle at the leading order and is 
given by, 
\bea
\bar{\theta}_{08}&=&-\frac{1}{2} 
\arctan\frac{2|M_{08}^2|}{M_{0}^2-M_{88}^2}. 
\eea
The self energy in Eq.(\ref{eq:self}) can be diagonalized with
the following transformation,
\bea
\begin{pmatrix} 
\eta_8 \\
\eta_0 \end{pmatrix}=\begin{pmatrix} \sqrt{z_{88}} & 0 \\ 0 & 1
\end{pmatrix} \begin{pmatrix} \cos \theta_{08} & \sin \theta_{08} \\
                              -\sin \theta_{08} & \cos \theta_{08} 
\end{pmatrix} \begin{pmatrix} \eta \\ \eta^\prime 
\end{pmatrix},
\label{eq:tr08}
\eea 
where,
\bea 
&&\begin{pmatrix}
m_{\eta}^2 & 0\\
0 & m_{\eta'}^2 \end{pmatrix}
=\nn \\
&&\begin{pmatrix} \cos \theta_{08} & -\sin \theta_{08} \\
                              \sin \theta_{08} & \cos \theta_{08} 
\end{pmatrix}
\begin{pmatrix}
M_{88}^2 z_{88} + \delta M^2_{88} & M^2_{08}\sqrt{z_{88}}+\delta M^2_{08}
\\ M^2_{08} \sqrt{z_{88}}+\delta M^2_{08} & M_{0}^2
\end{pmatrix}
\begin{pmatrix} \cos \theta_{08} & \sin \theta_{08} \\
                              -\sin \theta_{08} & \cos \theta_{08} 
\end{pmatrix}.
\label{eq:massmatrix}
\eea
We use the transformation Eq.(\ref{eq:tr08}) when we compute the form factors
for $\tau \to K \eta^{(\prime)} \nu$ decays. 
From Eq.(\ref{eq:massmatrix}), $M_{0}$ and $M_{08}$ are written by,
\begin{eqnarray}
M_{0}^2 &=& m_\eta^2 + m_{\eta'}^2 - M_{88}^2 z_{88}-\delta M^2_{88} , \nn  \\
\sqrt{z_{88}}M_{08}^2+\delta M_{08}^2 &=& -\sqrt{M_{00}^2 (M_{88}^2 z_{88}+ \delta M^2_{88})-m_{\eta}^2 m_{\eta'}^2} .
\label{eq:M0M08}
\end{eqnarray}
Eq.(\ref{eq:M0M08}) can be used to obtain the input values for $M_{0}$ and $M_{08}$ from the mass spectrum when the finite counterterms are given.
The mixing angle $\theta_{08}$ including the correction is also given by, 
\bea
\theta_{08}&=&-\frac{1}{2} 
\arctan\frac{2|M_{08}^2\sqrt{z_{88}}+\delta M^2_{08}|}{M_{0}^2-M_{88}^2 z_{88}-\delta M^2_{88}}. 
\eea
When we compute the form factors related to $\eta$ and $\eta^\prime$ in 
one-loop order, the mixing angle $\theta_{08}$ implies one-loop corrected one. 
The treatment is consistent with the rigorous one-loop computation and the 
difference is at two loop order.
\subsection{vector meson sector}
Now we turn to the vector meson sector of the Lagrangian.
The quantum corrections to the chiral Lagrangian with vector mesons
have been discussed in several works
\cite{Tanabashi:1993np, Rosell:2004mn, Harada:2003jx}.
They study the vector meson loop correction in addition to the loop correction
due to pseudoscalar mesons. 
Our aim is to construct the effective theory which can be used
to study the process in which a single vector meson can be nearly on-shell.
The corresponding energy region for hadronic invariant 
mass is $E_h < 2M_V$. For the hadronic $\tau$ decays, 
this approach is valid
in the energy region $E_h< 1400 \sim 1600 $(MeV).
In the region, the vector meson does not contribute to the loop diagram 
and only the light pseudoscalar mesons loop should be 
taken into account. The loops of the soft pseudoscalar mesons  can be systematically included using the momentum and loop expansion.  We regard
the vector meson as the classical background field. As for pseudoscalar 
mesons, we split them into hard classical background field and the soft 
quantum fluctuation.  For example, the decay products  $K \pi$ of $K^\ast$
meson decay have the hard momentum $E_h \sim M_V$ and they are 
treated as the classical background field.  
Though the vector mesons
do not contribute in the loop diagrams, they
contribute to the amplitude as intermediate dressed propagator
which connects
the 1 PI (Particle Irreducible) 
vertices of vector mesons and pseudoscalar mesons.
The self-energy of the vector meson and 1 PI vertices with or without vector
meson legs can be systematically improved by taking the quantum corrections
of the soft pseudoscalar meson loops into account.
1 PI vertices can be renormalized by adding the counterterms.
What kind of counterterms should be added depends on the
number of the pseudoscalar loops $N$ and the number of the external 
legs of the vector 
meson $N_V$.  
We focus on the chiral limit.
The number of the derivatives for the counterterms 
is determined by superficial degree of the divergence
of the 1 PI diagrams. As we prove later, 
the superficial degree of divergence of 1 PI diagram
of N loop order and $N_V$ external vector meson legs is
given as,
\bea
\omega=2N + 2-N_V. 
\label{eq:super}
\eea
This formula tells us the types of the counterterms which should
be added when we carry out N loop order computation. 
In general, the local counterterms and the finite counterterms 
can be classified with the number of derivative $n_d$
and the number of the vector mesons $n_V$ in the
Lagrangian. The interaction term with $n_d$ 
derivatives and with $n_V$ vector meson fields has the form of, 
\bea
{\cal L}=F(\xi)\partial^{n_d} V^{n_V}, 
\eea
where the Lorents indices are contracted appropriately.
$F(\xi)$ denotes some function of the chiral field. 
The derivatives can act on both the chiral field
and the vector field $V$.
Since the number of derivatives of the vertex of counterterms is equal to 
the superficial degree of divergence $\omega$,
the divergence of the N loop order Feynman diagram with $N_V$ external 
vector mesons 
can be subtracted by the counterterm
with the following number of the derivatives and the vector meson legs,
\bea
(n_d,n_V)=(2N+2-N_V, N_V).
\eea
In table I,
we show $(n_d, n_V)$ for a given set of $N$ and $N_V$. 
\begin{table} 
\begin{center}
\begin{tabular}{|c|c|c|c|c|c|} \hline
N & 0 & 1 & 2&  N \\ \hline
$n_d+n_V$   & $2$ &$ 4$ &$6$ & $2N+2$ \\ \hline 
$N_V$ &       &       &        &     \\  \hline 
$0$   & $(2,0)$&$(4,0)$ &$(6,0)$& $(2N+2,0)$ \\
$1$   & $(1,1)$&$(3,1)$ &$(5,1)$& $(2N+1,1)$ \\
$2$   & $(0,2)$ &$(2,2)$ &$(4,2)$& $(2N,2)$ \\
$3$   &         &$(1,3)$ &$(3,3)$& $(2N-1,3)$ \\
$4$   &         &$(0,4)$ &$(2,4)$& $(2N-2,4)$ \\
$5$   &         &      & $(1,5)$ &$(2N-3,5)$ \\
$6$   &         &      & $(0,6)$ &$(2N-4,6)$ \\
$.$   &         &      &         &   $..$  \\  
$2N+1$&         &      &         &$(1,2N+1)$ \\
$2N+2$&         &      &          &$(0,2N+2)$ \\ \hline 
\end{tabular}
\caption{The number of the derivatives $n_d$ and the number of the
vector meson external legs $n_V$.
$(n_d,n_V)$ are shown for a given set of numbers of loop $N$
and the number of the vector meson legs $N_V$.}  
\end{center}
\label{table:1}
\end{table}
The lowest order Lagrangian corresponds to $N=0$
case in the table I and it 
includes the interaction terms of the type,
\bea
(n_d, n_V)=(2,0),(1,1),
\eea
where $(n_d,n_V)=(1,1)$ corresponds to the term of ${\rm Tr}[V^\mu \alpha_\mu]$
in Eq.(\ref{eq:lag}).  The lowest order Lagrangian includes mass term of the
vector mesons while the kinetic term is not included. This is in contrast to
the approach of \cite{Tanabashi:1993np, Harada:2003jx} where the vector boson
is treated as gauge boson and the 
kinetic term is included in the leading order Lagrangian.
The vertices of only vector mesons do not contribute to vertices in 
any 1 PI loop
diagrams since the vector meson does not contribute in the loop diagram.
Such vertices include the tree level mass term with $(n_d,n_V)=(0,2)$
and  N loop order counterterms with $(n_d, n_V)=(0,2N+2)$ which has 
the form  $V^{2N+2}$.   
Now we prove Eq.(\ref{eq:super}).
We consider a 1 PI diagram of N loop order.
N loop order diagram
includes the N loop diagrams with the tree vertex as well as the
diagrams with higher loop order vertices with the number of loop $n_L$
less than $N$. 
We denote the number of 
n loop order type vertices with $n_d$ 
derivatives and $n_V$ vector meson fields  included in the diagram
as $n_v^{(n)}[n_d,n_V]$.
Note that $n_d+n_V=2n+2$. The total number of the n loop order 
interaction vertices in the 1 PI diagram is given by
\bea
N_v^{(n)}=\sum_{n_V=0}^{2n_V+1} n_v^{(n)}[2n+2-n_V,n_V].
\eea
Although in N loop order 1 PI diagram
consists of the various loop order vertices,
the number of the vertices must satisfy the following relation
\bea
N=n_L+ \sum^{N}_{n=1} n N_v^{(n)},
\label{eq:nL}
\eea
The number of pseudoscalar meson internal propagator
$I_B$ is written as,
\bea
I_B=\sum^N_{n=0} N_v^{(n)}+n_L-1.
\label{eq:IB}
\eea
Then one can compute the superficial divergence $\omega$ of the 1 PI diagram,
\bea
\omega=4n_L-2 I_B +\sum_{n=0}^N 
\sum_{n_V=0}^{2n+1}(2n+2-n_V) n_v^{(n)}[2n+2-n_V,n_V].
\label{eq:ome}
\eea
The last term of Eq.(\ref{eq:ome}) is the number of the
derivatives of the diagram. Substituting Eq.(\ref{eq:IB})
and Eq.(\ref{eq:nL}) into Eq.(\ref{eq:ome}),
one can show Eq.(\ref{eq:super}) as,
\bea
\omega&=&2 N+2 -\sum_{n=0}^N 
\sum_{n_V=1}^{2n+1} n_V n_v^{(n)}[2n+2-n_V,n_V], \nn \\
      &=& 2N+2-N_V.
\eea

In the same way as the chiral perturbation theory, we
rely on the momentum expansion. Because the loop momentum of
the pseudoscalar mesons is soft, the expansion is valid. 
In the Lagrangian, there is no kinetic term for the
vector meson at the leading order. The kinetic term is generated as the   
loop correction of the pseudoscalar mesons.
According to Eq.(\ref{eq:super}), one can extend the chiral counting 
to the case with vector mesons. 
In generalized chiral counting, the vector meson field $V_\mu$ is
counted as
$O(p)$ and the chiral breaking 
$\chi$ is counted as $O(p^2)$. 
The couterterms for $O(p^4)$ 
are obtained by computing divergent part of the one loop corrections
due to pseudoscalar mesons. We use the 
background field method and 
the corrections can be computed and the counterterms can be determined so that
they are consisitent with chiral symmetry \cite{Gasser:1984gg}.
The outline of the derivation is shown in Appendix A and they are given by,
\def\aTr{{\rm Tr}}
\bea
L_c&=&
K_1  2 i \aTr(\alpha_{\perp \mu} \alpha_{\perp \nu})
(D_\mu v_\nu-D_\nu v_\mu+i[v_\mu, v_\nu])\nn \\
&-&
\frac{1}{2} \left(K_2
\aTr(\xi^\dagger F_{L \mu \nu} \xi)(D_\mu v_\nu-D_\nu v_\mu+i[v_\mu, v_\nu])+
K_3 \aTr(D_\mu v_\nu-D_\nu v_\mu+i[v_\mu, v_\nu])^2 \right)
\nn \\
&+&\frac{4B}{f^2} \left(K_4
\aTr\{(\xi M \xi+ \xi^\dagger M \xi^\dagger) v^2\} +K_5
\aTr\{M(U+U^\dagger)\} \aTr(v^2) \right) \nn \\
&+& K_6 \aTr(v_\rho \alpha_\perp^\mu)  \aTr(v^\rho \alpha_{\perp \mu})+ 
K_7\aTr(v^2 \alpha_{\perp \mu} \alpha_{\perp}^\mu)+K_8
\aTr(v^2) \aTr(\alpha_{\perp \mu} \alpha_\perp^\mu) \nn \\
&+&K_9\{\aTr(v^2)\}^2+K_{10} \aTr(v^4) \nn \\  
&+&i  \eta_0 T_1 \frac{g_{2p}}{f^2} 
\aTr\{(\xi M \xi- \xi^\dagger M \xi^\dagger) v^2
\} 
+i \eta_0 T_2 \frac{g_{2p}}{f^2}
\aTr\{M(U-U^\dagger)\} \aTr(v^2) \nn \\
&+& T_3 \frac{g_{2p}}{f^2}
i \frac{4B}{f^2} {\eta_0}
\aTr M(U+U^\dagger) \aTr M(U-U^\dagger)
+T_4 \left(\frac{g_{2p}}{f^2} \right)^2 {\eta_0}^2 
\left(\aTr M(U-U^\dagger) \right)^2 \nn \\
&+&i T_5 \frac{4 B}{f^2}
\frac{g_{2p}}{f^2}\eta_0 \aTr(MUMU-MU^\dagger M U^\dagger)\nn \\
&+&
T_6 \left(\frac{g_{2p}}{f^2} \right)^2
\eta_0^2 \aTr(MUMU+MU^\dagger M U^\dagger-2 M^2) \nn \\
&+&i \frac{g_{2p}}{f^2}
\eta_0\bigl{[}T_7 \aTr\{M (D_{L\mu}U(D_{L}^\mu U)^\dagger U-U^\dagger
D_{L\mu}U(D_{L}^\mu U)^\dagger)\} \nn \\
&+&T_8
\aTr(M(U-U^\dagger)) \aTr(D_{L\mu}U(D_{L}^\mu U)^\dagger) \bigr{]}
\nn \\
&+&L_1 \{\aTr(D_{L\mu}U(D_{L}^\mu U)^\dagger)\}^2 +L_2  \aTr \{D_{L}^\mu U (D_L^\nu U)^\dagger\} 
\aTr \{D_{L\mu}U (D_{L \nu} U)^\dagger\} \nn \\
&+&L_3 \aTr \{D_{L}^\mu U (D_{L \mu} U)^\dagger D_{L}^\nu U 
(D_{L \nu} U)^\dagger\} \nn \\
&+& L_4 
\aTr(D_{L\mu}U(D_{L}^\mu U)^\dagger) \aTr\{M(U+U^\dagger)\} \frac{4B}{f^2}
+ L_5 \aTr\{D_{L\mu}U(D_{L}^\mu U)^\dagger 
(UM+MU^\dagger)\} \frac{4B}{f^2}
\nn \\
&+& \frac{16 B^2}{f^4}
\{L_6 \{\aTr M(U+U^\dagger)\}^2+L_7\{\aTr M(U-U^\dagger)\}^2\} \nn \\
&+&L_8 \frac{16 B^2}{f^4}
\aTr(MUMU+MU^\dagger MU^\dagger) +i L_9 
\aTr\{F_{L \mu \nu} D_L^{\mu} U (D_L^{\nu} U)^\dagger\} \nn \\
&+& H_1 \aTr F_{L \mu \nu} F_L^{\mu \nu} + H_2 
\left(\frac{4 B}{f^2}\right)^2 \aTr(M^2) + H_3 M_0^4,
\label{eq:fullcounter}
\eea
where $v_\mu=\frac{M_V^2}{2gf^2}(V_\mu-\frac{\alpha_\mu}{g})$.
The covariant derivative is defined as; 
$D_\mu v_\nu=\partial_\mu v_\nu + i [\alpha_\mu, v_\nu]$.
$\alpha_{\perp}$ in Eq.(\ref{eq:fullcounter}) is given as,
\bea
\alpha_{\perp \mu}&=&\frac{\xi^\dagger D_{L \mu} \xi-\xi \partial_\mu \xi^\dagger}{2i}.    
\eea
The coeffcients of the counterterms are splitted into the finite parts
and divergent parts as,
\bea
K_i&=&\lambda k_i +K^{r}_i (i=1 \sim 10), \nn \\
T_i&=& \lambda t_i +T^{r}_i (i=1 \sim 6), \nn \\
L_i&=&\lambda \Gamma_i +L^{r}_i (i=1 \sim 9), \nn \\
H_i&=&\lambda \Delta_i +H^{r}_i (i=1 \sim 3),
\label{eq:finitecounter}
\eea 
where, 
\bea
\lambda=-\frac{1}{32 \pi^2}(C_{UV}+1-\log \mu^2),
\eea
with $C_{UV}=\frac{1}{\epsilon}-\gamma+ \log 4 \pi$.
The coefficients $k_i, t_i, \Gamma_i$ and  $\Delta_i$ are given in the 
Table II.
\begin{table}
\begin{center}
\begin{tabular}{|c|c|c|c|} \hline
$k_1=1$ & $t_1=-6$ & $\Gamma_1=\frac{2c^2+1}{32}$ & $\Delta_1=-\frac{1}{8}$
\\  \hline
$k_2=1$& $t_2=-2$ & $\Gamma_2=\frac{1+2c^2}{16}$ & 
$\Delta_2=\frac{5}{24}+\frac{g_{2p}^2 f^2}{4 B^2}$
\\  \hline
$k_3=1$& $t_3=-\frac{11}{18}$& $\Gamma_3=\frac{3(c^2-1)}{16}$ & $\Delta_3=\frac{1}{2}$  \\ \hline
$k_4=\frac{3}{2}$& $t_4=-\frac{11}{9}$ & $\Gamma_4=\frac{c}{8}$ & \\ \hline 
$k_5=\frac{1}{2}$& $t_5=-\frac{5}{6}$ &  $ \Gamma_5=\frac{3c}{8}$ & \\ \hline
$k_6=4c$& $t_6=-\frac{5}{3}$ &   $\Gamma_6=\frac{11}{144}-\frac{g_{2p}^2 f^2}{24 B^2}$ & \\ \hline
$k_7=6c$& $t_7=-\frac{3c}{2}$  &   $\Gamma_7=0$ & \\ \hline
$k_8=2c$& $t_8=-\frac{c}{2}$ &   
$\Gamma_8=\frac{5}{48}+\frac{g_{2p}^2 f^2}{8 B^2}$ & \\ \hline 
$k_9=-3$& &  $\Gamma_9=\frac{1}{4}$ &\\ \hline
$k_{10}=-3$ & &  &\\ 
\hline  
\end{tabular}
\caption{The coefficients of the counterterms;
$k_i, \Gamma_i$ and $\Delta_i$. $c=1-\frac{M_V^2}{g^2 f^2}$.}
\end{center}
\label{table:2}
\end{table} 
From Eq.(\ref{eq:fullcounter}), we extract
the effective counterterms which are relevant for the calculation
of the form factor of $\tau \to K \pi \nu$ decay.
The effective counterterms which subtract the
divergence of the amplitudes which contains a vector meson in Fig.1,
can be deduced from the counterterms shown in Eq.(\ref{eq:fullcounter}).
They are the counterterms for the self energy of vector mesons, $V \to P P$ vertex,
and the production amplitude of the vector meson; $A_L \to V$ and are  
defined as,  
\bea
{\cal L}^{eff}_c&=&-\frac{1}{2}
Z_V {\rm Tr} (F^0_{V\mu \nu} F_{V}^{0 \mu \nu})
\nn \\
&+& C_1 {\rm Tr}\left[\frac{\xi \chi \xi + \xi^{\dagger} \chi^{\dagger}
 \xi^{\dagger}}{2}  (V_\mu-\frac{\alpha_{\mu}}{g})^2 \right] 
+ C_2 {\rm Tr}\left(\frac{\xi \chi \xi + \xi^{\dagger} \chi^{\dagger}
 \xi^{\dagger}}{2}\right)
 {\rm Tr}\left[(V_{\mu}-\frac{\alpha_{\mu}}{g})^2 \right] \nn \\
&+& i \frac{C_3}{f^2}  {\rm Tr} (F_V^{0 \mu \nu} 
\partial_{\mu} \pi \partial_\nu \pi)
+ C_4 {\rm Tr}( F_V^{0 \mu \nu} F^0_{L \mu \nu}),
\label{eq:counter}
\eea
where $F^0_{V\mu\nu} = \partial_\mu V_\nu - \partial_\nu V_\mu$ 
and $F^0_{L\mu\nu} = \partial_\mu A_{L \nu} 
- \partial_\nu A_{L\mu}$.. 
$Z_V$ and $C_i$ ($i=1,\ldots,4$) are 
renormalization constants.  
The coefficients $C_i$ can be written in terms of the coefficients
of the counterterms of Eq.(\ref{eq:fullcounter}),
\bea
Z_V&=& K_3 \left(\frac{M_V^2}{2 g f^2}\right)^2, \nn \\
C_1&=&2 K_4 \left(\frac{M_V^2}{2 g f^2}\right)^2, \nn \\
C_2&=&2 K_5 \left(\frac{M_V^2}{2 g f^2}\right)^2, \nn \\
C_3&=& \frac{M_V^2}{gf^2}(K_1-K_3 \frac{M_V^2}{2 g^2 f^2}), \nn \\
C_4&=&-\frac{M_V^2}{4gf^2}(K_2-K_3 \frac{M_V^2}{2 g^2 f^2}).
\label{eq:rel}
\eea
The finite parts of the counterterms also satisfy the relations 
similar to Eq.(\ref{eq:rel}),
\bea
Z^r_V&=& K^r_3 \left(\frac{M_V^2}{2 g f^2}\right)^2, \nn \\
C^r_1&=&2 K^r_4 \left(\frac{M_V^2}{2 g f^2}\right)^2, \nn \\
C^r_2&=&2 K^r_5 \left(\frac{M_V^2}{2 g f^2}\right)^2, \nn \\
C^r_3&=& \frac{M_V^2}{gf^2}(K^r_1-K^r_3 \frac{M_V^2}{2 g^2 f^2}), \nn \\
C^r_4&=&-\frac{M_V^2}{4gf^2}(K^r_2-K^r_3 \frac{M_V^2}{2 g^2 f^2}). 
\label{zrcr}
\eea
One can extract the counterterms for the $1 PI$ vertex of the type
$A_L \to P P$.
They are given as,
\bea
{\cal L}^c_{1 PI}&=&i\frac{C_5}{f^2}
{\rm Tr}{F^0_{L \mu \nu} 
\partial^\mu \pi \partial^\nu \pi} \nn \\
&+&i \Bigl\{-K_4 \frac{1}{2 f^2}
\left(\frac{M_V^2}{2 g^2 f^2}\right)^2+ 4  \frac{L_5}{f^2}\Bigr\}  
({\rm Tr} A_{L \mu} \{[\pi, \partial^\mu \pi], \chi \}) \nn
\\
&+& i \Bigl\{ 8 \frac{L_4}{f^2}- \frac{K_5}{f^2} 
\left(\frac{M_V^2}{2 g^2 f^2} \right)^2 \Bigr\}
{\rm Tr}A_{L \mu} [\pi, \partial^\mu \pi] {\rm Tr}\chi \nn \\
&+& \frac{4 L_5 i}{f^2} {\rm Tr} A_{L \mu} 
\{\partial^\mu \pi,[\pi, \chi] \},
\eea
where $C_5$ and its finite part $C_5^r$ are given by,
\bea
C^{(r)}_5=\frac{M_V^2}{2 g^2 f^2}(-K^{(r)}_1-K^{(r)}_2+\frac{M_V^2}{2 g^2 f^2}
K^{(r)}_3) +4 L^{(r)}_9. 
\label{c5}
\eea

We briefly comment on an intrinsic
parity violating interaction and its contribution to the 
vector meson self-energy.
After quarks are integrated out, 
the intrinsic parity violating interaction term
of two vector mesons and a pseudoscalar meson can be generated. 
One may wonder if there is some
contribution to the self-energy of $K^\ast$ meson
due to one loop diagram of a vector meson and a soft pion with the intrinsic parity violating 
vertex of $K^\ast \to  K^\ast \pi$ and $K^\ast \to \rho K$.
However, the vertex with a soft pseudoscalar meson is absent as the reason 
is given below.
 When the background field satisfies the equation motion
, the first variation 
with respect to the soft pion quantum fluctuation $\Delta$ vanishes in the interaction Lagrangian.
We have shown this in  Eq.(\ref{eq:soft}) and Eq.(\ref{eq:bfeq})
for the intrinsic parity conserving case. 
This conclusion does not change even the intrinsic parity
violating terms are included in the Lagrangian.
Then the vertex with two vector mesons and a soft pion is absent
and the vector meson loop with the soft pion does not contribute to the self-energy
of vector mesons. 
\section{The form factors at $O(p^4)$}
\begin{figure}[htbp]
\begin{center}
\resizebox{10cm}{!}{\includegraphics{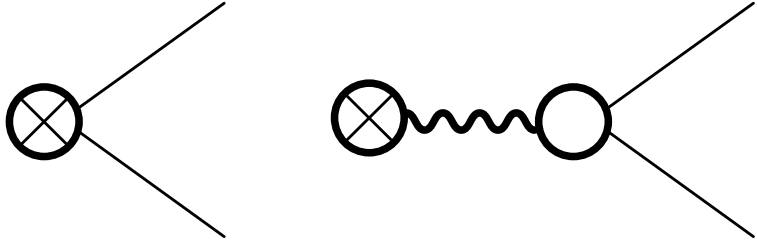}}
\caption{Feynman diagrams which 
contribute to the form factors. The diagrams are classified into two categories
, one of them corresponds to 1 particle irreducible diagram. The other
is the diagram with a vector meson propagator. 
The crossed circle
denotes the weak vertex. The circle denotes the interaction vertex of the vector meson and the two pseudo scalar mesons.
}
\end{center}
\label{fig:fig1}
\end{figure}
In this section, we compute the form factors. The matrix elements 
of the current $\bar{u} \gamma_\mu s$ are obtained by identifying
the quark current with the corresponding current of the chiral Lagrangian,
Eq.(\ref{eq:lag}) and Eq.(\ref{eq:fullcounter}).
In $ \tau \to K \pi \nu$ process, 
$K^*$ meson which is produced by the strangeness changing
current can contribute to the vector form factor.  
The resonant contribution is significant when $K \pi$ invariant
mass is near to the resonance pole. 
In our framework, the resonant contribution is included in the second
diagram of Fig.1.
We take into account the resonance contribution by using the vector meson propagator with  one-loop corrections to self energy. 
Since the propagator have a pole in complex plane, 
the effect of the width of resonance is also included.
Thus we can reproduce the resonance behavior.

There are three parts of the diagrams of Fig.1.
The first one is 1 particle irreducible (1 PI) diagrams 
and the diagrams with one loop corrections are shown in Fig.2. 
They correspond to the one loop 
corrections to $W^+ \to K^+  \pi^0$ vertex.
They include all the diagrams which are also present in
chiral perturbation within one loop.
Their contributions to the matrix element $\langle K^+ \pi^0|\overline{u_L}\gamma_\mu s_L |0 \rangle$ become,
\bea
&& \langle K^+ \pi^0|\overline{u_L}\gamma_\mu s_L |0 \rangle
\bigr|_{\rm 1 PI} = \nn \\
&& -\frac{c}{16 \sqrt{2} f^2}
\left[ Q_\mu (3 I_{\eta_8} + 2 I_K -5 I_\pi)
+2 q_\mu (3 I_{\eta_8} + 8 I_K + 7 I_\pi)
\right] \nn \\
&& + \frac{1}{16 \sqrt{2}f^2}\left(1-\frac{M_V^2}{2 g^2 f^2}
\right)
\left[\chi_\mu^{\pi K}\{(3-5c) \Sigma_{K \pi} +5 c Q^2\}
+ \chi_\mu^{\eta_8 K} \{(-1+3c) \Sigma_{K \pi} -3 c Q^2\}
\right] \nn \\
&& + \frac{3}{16 \sqrt{2}f^2}\left(1-\frac{M_V^2}{2 g^2 f^2} \right)^2
(J_\mu^{\pi K}+ J_\mu^{\eta_8 K})-\frac{1}{2 \sqrt{2}}\left(1-\frac{M_V^2}{2g^2 f^2}\right) q_\mu (\sqrt{z_K z_\pi}-1) \nn \\
&& + {\rm counterterms},
\label{eq:ff1}
\eea
 where $Q=p_K+p_\pi$ and $q=p_K-p_\pi$. 
$\Delta_{PQ}$ denotes the mass squared difference $\Delta_{PQ}=m_P^2-m_Q^2$.
$\Sigma_{PQ}$ denotes the sum of the mass squared $\Sigma_{PQ}=m_P^2+m_Q^2$.
The loop functions $I_P, \chi_\mu^{QP}, J_\mu^{QP}$ are 
given in Eq.(\ref{eq:loopf1}) and Eq.(\ref{eq:loopf2}).
$z_K$ and $z_\pi$ are finite wave function renormalization and they are 
given as,
\bea
z_K&=&1+3 c(\mu_K+\frac{1}{2} (\mu_{\eta_8} +\mu_\pi))-\frac{8}{f^2}\{L_4^{r}(2 m_K^2+m_\pi^2)+L_5^{r} m_K^2 \}, \nn \\
z_\pi&=&1+c(2 \mu_K+ 4 \mu_\pi)-
\frac{8}{f^2}\{ L_4^{r}(2 m_K^2+m_\pi^2)+L_5^{r} m_\pi^2 \}.
\label{eq:zkp}
\eea
Including the finite part of the counter terms, the result of the
1 PI part is
\bea
\langle K^+ \pi^0|\bar{u_L} \gamma_\mu s_L|0 \rangle
\Bigr{|}_{1 PI}
&=&
\frac{1}{2\sqrt{2}}(q_\mu-\frac{\Delta_{K \pi}}{Q^2} Q_\mu)
\Bigl{[}
-\frac{3 c}{2} (H_{K \pi}+H_{K \eta_8}) 
+\frac{c M_V^2}{8 g^2 f^2}
(10 \mu_K + 3 \mu_{\eta_8} + 11 \mu_\pi)\nn \\
&-&\frac{3}{8}
\left(\frac{M_V^2}{g^2 f^2}\right)^2 (H_{K \pi}+H_{K \eta_8}+\frac{2\mu_K+\mu_\pi+\mu_{\eta_8}}{2})-\frac{C_5^{r}}{2} \frac{Q^2}{f^2} +\nn \\
&& \frac{M_V^2}{2 g^2 f^2} \Bigl\{ \frac{M_V^2}{2 g^2 f^2}K_4^{r} \frac{m_K^2}{f^2}- 4 L_5^{r} \frac{\Sigma_{K \pi}}{f^2}+\frac{2 m_K^2+m_\pi^2}{f^2} (\frac{M_V^2}{2 g^2 f^2} K_5^{r}-8 L_4^{r})
\Bigr\}  \Bigr{]}\nn \\ 
&+&\frac{1}{2 \sqrt{2}} \frac{Q_\mu}{Q^2} \times \nn \\
&& \Bigl{[}(1-\frac{M_V^2}{2g^2f^2})\Bigl\{
-\frac{\Delta_{K \pi}\bar{J}_{K \pi}}{8f^2}\{
5 c Q^2-(5c-3) \Sigma_{K \pi}\}  \nn \\
&+&\frac{\Delta_{K \eta}\bar{J}_{K \eta} \cos^2 \theta_{08}+ 
\Delta_{K \eta^\prime} \bar{J}_{K \eta^\prime} 
\sin^2 \theta_{08}}{8f^2} 
\{3c Q^2 -(3c-1) \Sigma_{K \pi}\} \Bigr\} \nn \\
&+&\frac{3 \Delta_{K \pi}}{8f^2}
(1-\frac{M_V^2}{2g^2f^2})^2
\{ \frac{\Delta_{K \pi}^2}{s} \bar{J}_{K \pi}
+\frac{ \Delta_{K \eta}^2}{s} \bar{J}_{K \eta}\cos^2 \theta_{08}
+\frac{ \Delta_{K \eta^\prime}^2}{s} \bar{J}_{K \eta^\prime}\sin^2 \theta_{08}
\} \Bigr{]}
\nn \\
&+&\frac{1}{2 \sqrt{2}} \frac{\Delta_{K \pi}Q_\mu}{Q^2} 
\Bigl{[} \frac{c}{4} Q^2 \frac{3 \mu_{\eta_8}+2 \mu_K-5 \mu_\pi}{\Delta_{K \pi}}
+c \frac{M_V^2}{8 g^2 f^2}(10 \mu_K +3 \mu_{\eta_8}+11\mu_\pi)\nn \\
&-&\frac{3}{16}
\left(\frac{M_V^2}{g^2 f^2}\right)^2(2 \mu_K+\mu_\pi+\mu_{\eta_8}) -4 L_5^{r} 
\frac{Q^2}{f^2} +
\frac{M_V^2}{2 g^2 f^2} \times \nn \\
&&\Bigl\{ \frac{M_V^2}{2 g^2 f^2}K_4^{r} \frac{m_K^2}{f^2}- 4 L_5^{r} \frac{\Sigma_{K \pi}}{f^2}+\frac{2 m_K^2+m_\pi^2}{f^2} (\frac{M_V^2}{2 g^2 f^2} K_5^{r}-8 L_4^{r})
\Bigr\}
\Bigr{]}.
\eea
$L_i^{r}$ and $C^{r}_i$ are finite parts of the counterterms
$L_i$ and $C_i$ respectively.
The function $H_{PQ}$ is written in terms of the functions defined
in Eq.(\ref{eq:ML}) as,
\bea
H_{PQ}=\frac{1}{f^2}(Q^2 M_{PQ}^{r}-L_{PQ}).
\label{eq:HPQ}
\eea
$\bar{J}_{PQ}$ can be found in Eq.(\ref{eq:Jbar}) and Eq.(\ref{eq:Jbar2}).
We also introduce the following notations in this paper. 
\begin{eqnarray*}
Y_{K\eta_8} &=& Y_{K\eta}\cos^2\theta_{08} + Y_{K\eta'}\sin^2\theta_{08},
\end{eqnarray*}
where $Y=H, M^r$ and $L$ which also appear in the following equations.
\begin{figure}[thbp]
\resizebox{10cm}{!}{\includegraphics{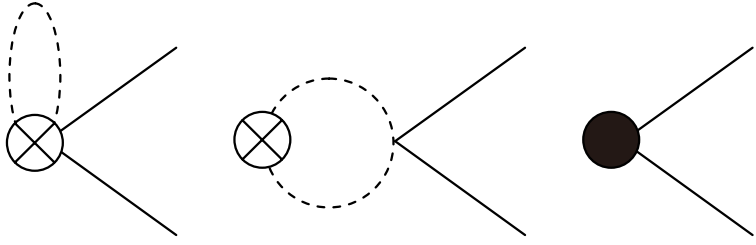}}
\caption{
The one loop 1 particle irreducible Feynman diagrams contributing 
$W^+ \to K^+ \pi^0$
form factors. The counterterms are shown by the black blob vertex.}
\label{fig:fig2}
\end{figure}
The diagram with a vector meson propagator is shown
in Fig.1.  It includes the diagram
with a $K^\ast$ propagator, $K^* \to K^+ \pi^0$ vertex and $W^+ \to K^\ast$ production amplitude.  The self-energy of the propagator, the vertex and the production amplitudes include one-loop corrections. 
We first compute the one loop corrections
to $K^\ast \to K \pi$ vertex which are shown in
Fig.3. 
\begin{figure}[bthp]
\resizebox{12cm}{!}{\includegraphics{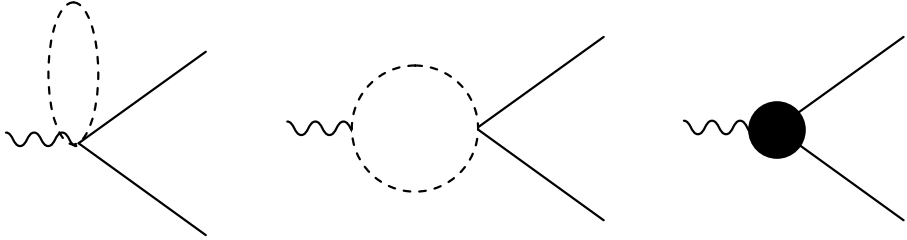}}
\caption{The one loop Feynman diagrams contributing to 
$K^{\ast +} \to K^+ \pi^0$ vertex. The black blob denotes the 
counterterm.} 
\label{fig:fig3}
\end{figure}
\bea
T_\mu(K^{\ast +} \to K^+ \pi^0)&=&-\frac{3 M_V^2}{32 gf^4}
(1-\frac{M_V^2}{2g^2f^2})(J_\mu^{\pi K}+J_\mu^{\eta_8 K})
-\frac{M_V^2 \Sigma_{K \pi}}{8gf^4}(\chi_\mu^{\pi K}-\frac{1}{4}\chi_\mu^{\eta_8 K}) \nn \\
&-& \frac{c M_V^2}{8gf^4}(Q^2-\Sigma_{K \pi})(
\chi_\mu^{\pi K}-\frac{3}{4}\chi_\mu^{\eta_8 K}
)+ \frac{3M_V^2}{32 gf^4} q_\mu(I_\pi+I_{\eta_8}+2 I_K)\nn \\
&+& \frac{M_V^2}{4 gf^2}
q_\mu (\sqrt{z_K z_\pi}-1)+ {\rm counter terms} \nn \\
&=&-\frac{3M_V^2}{8gf^4}(1-\frac{M_V^2}{2g^2f^2})
\{(M^r_{K \pi}+M^r_{K \eta_8})(Q_\mu \Delta_{K \pi}-Q^2 q_\mu)
+(L_{K \pi}+L_{K \eta_8}) q_\mu \} \nn \\
&+& \frac{M_V^2}{16 gf^2}q_\mu \Bigl\{-\frac{3 M_V^2}{g^2 f^2} \frac{2\mu_K+\mu_\pi+\mu_\eta}{2}+c(10\mu_K+3\mu_{\eta_8}+11\mu_\pi) \nn \\
&-&32 L_4^{r} \frac{2m_K^2+m_\pi^2}{f^2}
-16 L_5^{r} \frac{\Sigma_{K \pi}}{f^2}
\Bigr\} \nn \\
&+&\frac{q_\mu}{4gf^2}\{C_2^{r}(2 m_K^2+m_\pi^2)+C_1^{r}m_K^2\}+ \frac{C_3^{r}}{8 f^2}(Q^2 q_\mu-\Delta_{K \pi}Q_\mu) \nn \\
&+&  
\frac{M_V^2}{8 gf^4}\frac{Q_\mu}{s}\{\Sigma_{K \pi}(\Delta_{K \pi}\bar{J}_{K \pi}-\frac{1}{4}\Delta_{K \eta_8} \bar{J}_{K \eta_8}) \nn \\
&+&c(Q^2-\Sigma_{K \pi})
(\Delta_{K \pi}\bar{J}_{K \pi}-\frac{3}{4}\Delta_{K \eta_8} \bar{J}_{K \eta_8})\} 
\nn \\
&=& \Bigl\{-\frac{g}{2M_V^2}(1-\frac{M_V^2}{2g^2f^2})
(\delta B_{K^\ast}-Z^r_V)-\frac{C_3^{r}}{8f^2} \Bigr\} 
(Q_\mu\Delta_{K \pi}-Q^2 q_\mu)\nn \\
&-& q_\mu\frac{g}{2M_V^2} \Bigl\{(\delta A_{K^\ast}+Q^2 \delta B_{K^\ast})
(1-\frac{M_V^2}{2g^2f^2})
-C_1^{r} m_K^2-C_2^{r}(2 m_K^2+m_\pi^2) \Bigr\}
 \nn \\
&+& \frac{M_V^2}{16g f^2} q_\mu 
\Bigl\{-(2 \mu_K+\mu_\pi+\mu_{\eta_8})+c(10\mu_K+3\mu_{\eta_8}
+11\mu_\pi) \nn \\
&-&32 L_4^{r} \frac{2m_K^2+m_\pi^2}{f^2}
-16 L_5^{r} \frac{\Sigma_{K \pi}}{f^2}
\Bigr\}\nn \\
&+&\frac{M_V^2}{8g f^4}\frac{Q_\mu}{s}
\{\Sigma_{K \pi}(
\Delta_{K \pi} \bar{J}_{K \pi}-
\frac{1}{4} (\Delta_{K \eta} \bar{J}_{K \eta}
 \cos^2 \theta_{08}+\Delta_{K \eta^\prime} \bar{J}_{K \eta^\prime}
\sin^2 \theta_{08}) )\nn \\
&+&c(Q^2-\Sigma_{K \pi}) (\Delta_{K \pi} \bar{J}_{K \pi}- \frac{3}{4} 
(\Delta_{K \eta} \bar{J}_{K \eta}\cos^2 \theta_{08}+\Delta_{K \eta^\prime} \bar{J}_{K \eta^\prime}
\sin^2 \theta_{08})) \}, \nn \\
\eea
where $\Delta_{K \eta_8} 
\bar{J}_{K \eta_8}\equiv
\Delta_{K \eta} \bar{J}_{K \eta} \cos^2 \theta_{08}+ \Delta_{K \eta^\prime} 
\bar{J}_{K \eta^\prime} \sin^2 \theta_{08}$. 
Next, the propagator of the $K^\ast$ meson
is obtained by including one loop self energy
corrections. Using  Eq.(\ref{eq:invp}), the $K^\ast$ meson propagator is given by  $i D_{\mu \rho}$
where $D_{\mu \rho}$ is given by,
\bea
D_{\mu \rho}=\frac{g_{\mu \rho}-\frac{Q_\mu Q_\rho \delta B}{M_V^2+\delta A + Q^2 \delta B}}{M_V^2+\delta A},
\label{prop}
\eea
where the self energy corrections $\delta A$ and $\delta B$ in this section
are identical to
the $K^\ast$ mesons ones given in Eq.(\ref{del_ks}),
\bea
\delta A=\delta A_{K^\ast}, \quad \delta B=\delta B_{K^\ast}.
\eea
\begin{figure}[thbp]
\resizebox{10cm}{!}{\includegraphics{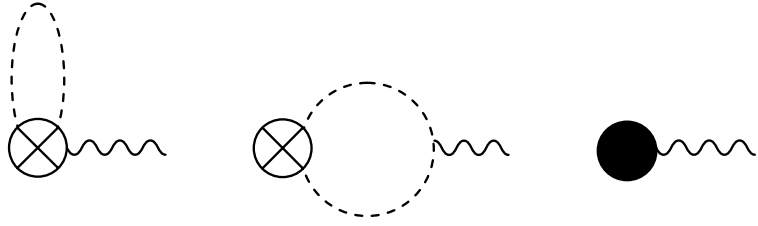}}
\caption{The one loop Feynman diagrams contributing 
$W^+ \to K^\ast$ production amplitude.}
\label{fig:fig4}
\end{figure}
The $K^\ast$ production amplitude with the one loop corrections 
are shown in Fig.4 and is given by,
\bea
\langle K^{\ast+}_\nu|\overline{u_L} \gamma_\mu s_L|0 \rangle  
&=&-\frac{3 M_V^2}{8g f^2 \sqrt{2}}\left(1-\frac{M_V^2}{2 g^2 f^2}\right)
(J_{\mu \nu}^{\pi K}+J_{\mu \nu}^{\eta_8 K}) \nn \\
&&+\frac{1}{\sqrt{2}g}[C_1 m_K^2+ C_2 (2m_K^2+m_\pi^2)] g_{\mu \nu}
+\sqrt{2} C_4 (Q_\mu Q_\nu -Q^2 g_{\mu \nu})
\nn \\
&&+ \frac{3 M_V^2}{8 \sqrt{2} g f^2}(I_\pi+I_{\eta_8}+2 I_K) 
g_{\mu \nu} \nn \\
&=&\frac{1}{\sqrt{2} g}
\left[ \left\{ \delta A+Q^2 \delta B-\frac{3 M_V^2}{2 f^2}
(L_{K \pi}+L_{K \eta_8})\right\} 
g_{\mu \nu} \right. \nn \\
&&+ \left.
\left\{Z^r_V-2g C^r_4-\delta B+\frac{3 M_V^2}{2 f^2}
(M^r_{K \pi}+M^r_{K \eta_8}) \right\}
(Q^2 g_{\mu \nu}-Q_\mu Q_\nu) \right],\nn \\
\label{eq:Kstar}
\eea
where,
\bea
C_4=\frac{M_V^2}{128 \pi^2 g f^2} \left(1-\frac{M_V^2}{2 g^2 f^2} \right)
(C_{UV}+1-\ln \mu^2)+C^r_4.
\eea
$J_{\mu \nu}^{QP}$ is defined as,
\bea
J_{\mu \nu}^{QP}&=&(g_{\mu \nu}-\frac{Q_\mu Q_\nu}{Q^2}) \left(-4f^2H_{PQ}+\frac{2}{3} \lambda Q^2-2\lambda \Sigma_{PQ}-2(\mu_Q+\mu_P)f^2 \right) \nn \\
&+& \frac{Q_\mu Q_\nu}{Q^2}\left(\frac{\Delta_{PQ}^2 \bar{J}_{PQ}}{s}-2\lambda \Sigma_{PQ}-2(\mu_Q+\mu_P) f^2 \right).
\eea
$J_{\mu \nu}^{\eta_8 K}$ is defined as;
$J_{\mu \nu}^{\eta_{8} K}=J_{\mu \nu}^{\eta K} \cos^2 \theta_{08}+
J_{\mu \nu}^{\eta^\prime K} \sin^2 \theta_{08}
$. 
Now one can assemble the contribution from the diagram with
a $K^\ast$ propagator to the form factor.
One can write $K^\ast$ production amplitude of the weak
vertex and $K^\ast \to K \pi$ decay amplitudes
as,
\bea
&& \langle K_\nu^{\ast} |\overline{u_L} \gamma_\mu s_L|0 \rangle
=g_{\nu \mu} G + (Q^2 g_{\nu \mu}-Q_\nu Q_\mu) {\cal H}, \nn \\
&&T_\rho (K^{\ast +} \to K^+ \pi^0)=E q_\rho + {\cal F}
 Q_\rho\Delta_{K \pi}, \nn \\
\eea
where $G,{\cal H},,E$ and ${\cal F}$ are given as,
\bea
G&=&
\frac{1}{\sqrt{2}g}\{M_V^2+\delta A+Q^2 \delta B-\frac{3 M_V^2}{2f^2}(L_{K \pi}+L_{K \eta_8})\}, \\
{\cal H}&=& \frac{1}{\sqrt{2}g}\{Z^r_V-2 g C^{r}_4-\delta B+ \frac{3 M_V^2}{2f^2}(M^{r}_{K \pi}
+M^{r}_{K \eta_8})\},\\
E&=&\frac{M_V^2}{4 gf^2}-\frac{g}{2 M_V^2}\{
(\delta A+ Q^2 \delta B)(1-\frac{M_V^2}{2 g^2 f^2})
-C_1^{r}m_K^2-C_2^r(2 m_K^2+m_\pi^2) \} \nn \\
 &+& \frac{M_V^2}{16 gf^2}\{-3 (2 \mu_K+\mu_\pi+\mu_{\eta_8})
+c(10 \mu_K+3\mu_{\eta_8}+11 \mu_\pi)-32 L_4^r\frac{2 m_K^2+m_\pi^2}{f^2}
-16 L_5^r \frac{\Sigma_{K \pi}}{f^2} \} \nn \\
&+& \{ \frac{g}{2 M_V^2}(1-\frac{M_V^2}{2 g^2 f^2})(\delta B-Z^r_V)
+\frac{C_3^{r}}{8f^2} \} Q^2, \\
{\cal F}&=&-\{ \frac{g}{2 M_V^2}(1-\frac{M_V^2}{2 g^2 f^2})(\delta B-Z^r_V)
+\frac{C_3^{r}}{8f^2} \} \nn \\
        &+& \frac{M_V^2}{8 g f^4}\frac{1}{Q^2}
\{ \Sigma_{K \pi}(\bar{J}_{K \pi}-
\frac{\bar{J}_{K \eta} \Delta_{K \eta}\cos^2 \theta_{08}+\bar{J}_{K \eta^\prime} \Delta_{K \eta^\prime}\sin^2 \theta_{08}}{4 
\Delta_{K \pi}}) \nn \\
&+&c (Q^2-\Sigma_{K \pi})(\bar{J}_{K \pi}-
\frac{3 (\bar{J}_{K \eta} \Delta_{K \eta} \cos^2 \theta_{08}+
\bar{J}_{K \eta^\prime} \Delta_{K \eta^\prime}\sin^2 \theta_{08})}
{4 \Delta_{K \pi}})
\}.
\eea
Using the form factors, we obtain,
\bea
&&\langle K^+ \pi^0|\overline{u_L} \gamma_\mu s_L|0 \rangle \Bigr{|}_{K^\ast}=
i (E q_\rho+F Q_\rho \Delta_{K \pi}) i D^{\rho \sigma} \left(g_{\sigma \mu}G+(Q^2 g_{\sigma \mu}-Q_{\sigma} Q_\mu) {\cal H} \right) \nn \\
&=&
-E \frac{G+Q^2 {\cal H}}{M_V^2+\delta A} 
(q_\mu-\frac{\Delta_{K \pi}}{Q^2} Q_\mu)
-G \frac{\Delta_{K \pi}}{Q^2}Q_\mu 
\frac{E+ Q^2 \cal{F}}{M_V^2+\delta A+ Q^2 \delta B}.
\eea
The vector form factor and the scalar form factors are defined as,
\bea
\langle K^+ \pi^0|\overline{u_L} \gamma_\mu s_L|0 \rangle
=\frac{1}{2} \{F_V (q_\mu-\frac{\Delta_{K \pi}}{Q^2} Q_\mu)+ F_S Q_\mu \}.
\eea
Then the contribution to the form factors is given as,
\bea
F_V^{K\pi}&=&F_V^{1 PI}+ F_V^{K^\ast}, 
\label{F_V}\\
F_S^{K\pi}&=&F_S^{1 PI}+ F_S^{K^\ast},
\label{F_S}
\eea
where,
\bea
F_V^{1 PI}&=&-\frac{1}{\sqrt{2}}(1-\frac{M_V^2}{2 g^2 f^2})+\nn \\
          &+&\frac{1}{\sqrt{2}}
\Bigl{[}
-\frac{3 c}{2} (H_{K \pi}+H_{K \eta_8}) 
+\frac{c M_V^2}{8 g^2 f^2}
(10 \mu_K + 3 \mu_{\eta_8} + 11 \mu_\pi)\nn \\
&-&\frac{3}{8}
\left(\frac{M_V^2}{g^2 f^2}\right)^2 (H_{K \pi}+H_{K \eta_8}+\frac{2\mu_K+\mu_\pi+\mu_{\eta_8}}{2})-\frac{C_5^{r}}{2} \frac{Q^2}{f^2} +\nn \\
&& \frac{M_V^2}{2 g^2 f^2} \Bigl\{ \frac{M_V^2}{2 g^2 f^2}K_4^{r} \frac{m_K^2}{f^2}- 4 L_5^{r} \frac{\Sigma_{K \pi}}{f^2}+\frac{2 m_K^2+m_\pi^2}{f^2} (\frac{M_V^2}{2 g^2 f^2} K_5^{r}-8 L_4^{r})
\Bigr\}  \Bigr{]}, \\ 
F_V^{K^\ast}&=& -2 E \frac{G + Q^2 {\cal H}}{M_V^2+\delta A} 
\label{FKs_Vexact} \\
&\simeq&-\frac{1}{2\sqrt{2}g} \frac{M_V^2}{M_V^2+\delta A}
 \left[ 4E + \sqrt{2}\frac{G+Q^2{\cal H}}{f^2}
 - \frac{M_V^2}{gf^2}
\right] , 
\label{FKs_V}\\
F_S^{1 PI}&=&\frac{1}{\sqrt{2}} \frac{1}{Q^2}\times \nn \\
&& \Bigl{[}(1-\frac{M_V^2}{2g^2f^2})
\Bigl\{-\frac{\Delta_{K \pi} \bar{J}_{K \pi}}{8f^2}\{
5 c Q^2-(5c-3) \Sigma_{K \pi}\}
\nn \\
&+&\frac{\Delta_{K \eta} \bar{J}_{K \eta}
\cos^2 \theta_{08} +\Delta_{K \eta^\prime} \bar{J}_{K \eta^\prime}
\sin^2 \theta_{08}}{8f^2}
\{3c Q^2 -(3c-1) \Sigma_{K \pi}\} \Bigr\} \nn \\
&&+\frac{3 \Delta_{K \pi}}{8f^2}(1-\frac{M_V^2}{2g^2f^2})^2 
\{\frac{\Delta_{K \pi}^2}{s} \bar{J}_{K \pi} 
+\frac{\Delta_{K \eta}^2}{s} \cos^2 \theta_{08}
\bar{J}_{K \eta}+\frac{\Delta_{K \eta^\prime}^2}{s} \sin^2 \theta_{08}
\bar{J}_{K \eta^\prime} \}\Bigr{]} \nn \\
&+&\frac{1}{\sqrt{2}}\frac{\Delta_{K \pi}}{Q^2}
\Bigl{[}-(1-\frac{M_V^2}{2 g^2 f^2})+\frac{c}{4} Q^2 \frac{3 \mu_{\eta_8}+2 \mu_K-5 \mu_\pi}{\Delta_{K \pi}}
+c \frac{M_V^2}{8 g^2 f^2}(10 \mu_K +3 \mu_8+11\mu_\pi)\nn \\
&-&\frac{3}{16}
\left(\frac{M_V^2}{g^2 f^2}\right)^2(2 \mu_K+\mu_\pi+\mu_{\eta_8}) -4 L_5^{r} 
\frac{Q^2}{f^2} +
\frac{M_V^2}{2 g^2 f^2} \times \nn \\
&&\Bigl\{ \frac{M_V^2}{2 g^2 f^2}K_4^{r} \frac{m_K^2}{f^2}- 4 L_5^{r} \frac{\Sigma_{K \pi}}{f^2}+\frac{2 m_K^2+m_\pi^2}{f^2} (\frac{M_V^2}{2 g^2 f^2} K_5^{r}-8 L_4^{r})
\Bigr\} \Bigr{]},\\
F_S^{K^\ast}&=& -2 G \frac{\Delta_{K \pi}}{Q^2}
 \frac{E+Q^2 {\cal F}}{M_V^2+\delta A+ Q^2\delta B} \label{FKs_Sexact}\\
&\simeq& -\frac{1}{2\sqrt{2}g}\frac{\Delta_{K \pi}}{Q^2} 
\frac{M_V^2}{M_V^2+\delta A+ Q^2\delta B}
\left[ 4(E+Q^2{\cal F}) + \sqrt{2}\frac{G}{f^2}
-\frac{M_V^2}{gf^2}
\right].
\label{FKs_S}
\eea
  For numerical calculation, we use  Eq.(\ref{FKs_V}) and Eq.(\ref{FKs_S})
which are obtained by omitting the two loop order contribution in the
numerators of Eq.(\ref{FKs_Vexact}) and Eq.(\ref{FKs_Sexact}).
To compare our form factors with the ones obtained by other methods,
we show the vector form factors in the chiral limit,
\bea
F_V^{1PI}&=&-\frac{1}{\sqrt{2}}\Bigl{[}(1-\frac{M_V^2}{2g^2f^2})
\{1+3(1-\frac{M_V^2}{2g^2f^2})H \}+\frac{C^{r}_5 Q^2}{f^2}) \Bigr{]},
\nn \\
F_V^{K^\ast}&=&-\frac{M_V^2}{2 \sqrt{2} g^2 f^2} \frac{M_V^2}{M_V^2+\delta A}
\left(1+6(1-\frac{M_V^2}{2g^2f^2})H +\frac{g (C^{r}_3-4C^{r}_4) Q^2}{2 M_V^2}\right).
\eea
The self energy correction of vector meson,
$\delta A$ in the chiral limit can be obtained with Eq.(\ref{del_rho}),
\bea
\delta A \rightarrow - Q^2 Z_V^{(r)}-3 M_V^2 H, 
\eea 
where $H$ is given by taking the chiral limit of $\frac{Q^2 M^{r}}{f^2}$ in 
Eq.(\ref{eq:M})
as,
\bea
H=\frac{Q^2}{12 f^2}[-\frac{1}{16 \pi^2}\log \frac{Q^2}{\mu^2}
+\frac{5}{48 \pi^2} + i \frac{1}{16 \pi}].
\eea
To compare our result with those of the other methods, we examine the case that
the vector meson dominace (VMD) relation 
$M_V^2=2 g^2 f^2$ holds. 
Then the vector form factor is written as,
\bea
F_V\Large{|}_{\rm chiral \ limit}^{VMD.}=-\frac{1}{\sqrt{2}}
\Bigr{[}\frac{M_V^2}{M_V^2-3 H M_V^2-Z_V^{r}Q^2}\{1+ g(C_3^r-4C_4^r)\frac{Q^2}{2 M_V^2}\}
+ \frac{C_5^r Q^2}{f^2}\Bigl{]}. 
\label{eq:chiral}
\eea
The result can be compared with the same limit of the form factor in \cite{Jamin:2006tk},
\bea
F_+=\frac{M_{K^\ast}^2 
e^{3{\rm Re.}(H)}}
{M_{K^\ast}^2-Q^2-iM_{K^\ast} \Gamma_{K^\ast}(Q^2)}.  
\eea
The difference of the overall factor $-\frac{1}{\sqrt{2}}$ is just due to the
the definition of the form factors. We observe that in the form factor of 
\cite{Jamin:2006tk}
, the chiral loop correction denoted by $H$ is exponentiated
and appears in the numerator of vector meson propagator while in our approach with the vector
dominance assumption, the chiral correction appears in the self-energy function in 
the denominator of the vector meson propagator. We also note that the finite counterterms 
generate linear $Q^2$ dependence in the form factor.  
They include the wave function renormalization constant of the vector meson $Z_V^{(r)}$,
and the other coefficients of the finite counter terms; 
$C_3^{r}$, $C_4^r$ and $C_5^r$. 
One can also compare our result with that of the resonance chiral theory
\cite{Rosell:2004mn}.
A difference of the form factor in \cite{Rosell:2004mn} 
from  Eq.(\ref{eq:chiral}) of our result
is that
the one loop corrections to their form factor depends quadratically on momentum squared
$Q^4$. This is due to the second derivatives coupling of the vector meson
to two pseudoscalars in their anti-symmetric tensor formulation of vector mesons.
In contrast to their approach,
the vector meson coupling into two pseudoscalar
meson coupling includes the first derivative. Therefore, the form factor of the
present approach depends on $Q^2$ linearly.
They also consider the loop contribution of all the resonances while in
our approach, the vector mesons do not contribute in the loop.
\section{Numerical analysis in the SM}
To evaluate the vector and scalar form factors, we fix 
{\bf $g, M_V$} and the coefficients of the counterterms by 
using the decay constants, masses and widthes of the mesons.
We also use the hadronic mass spectrum.
There are ten parameters, $\{g, M_V, Z_V^r, C_i^r, 
L_4^r, L_5^r \}$, $(i=1,\cdots,5)$ to be fixed.

From the matrix elements of the axial currents, we obtain the
pion and kaon decay constants \cite{Gasser:1984ux},
\begin{eqnarray}
f_\pi &=& f \left\{ 1 - c (2\mu_\pi+\mu_K)
 + 4 \left(\frac{m_\pi^2+2m_K^2}{f^2}L_4^r + \frac{m_\pi^2}{f^2}L_5^r
\right) \right\} ,
\label{fpi} \\
f_K &=& f \left\{ 1 - \frac{3c}{4} (\mu_\pi+2\mu_K+\mu_{\eta_8})
 + 4 \left(\frac{m_\pi^2+2m_K^2}{f^2}L_4^r + \frac{m_K^2}{f^2}L_5^r
\right) \right\} .
\label{fK}
\end{eqnarray}
Using the ratio of $f_K/f_\pi$,
we can write $L_5^r$ as follows,
\begin{eqnarray}
L_5^r = \frac{f^2}{4\Delta_{K\pi}} \left\{
 \frac{f_K}{f_\pi} -1 -\frac{c}{4} (5\mu_\pi-2\mu_K-3\mu_{\eta_8}) 
 \right\} .
\end{eqnarray} 
If we assume $f=f_\pi$, from Eq. (\ref{fpi}) $L_4^r$ is expressed
as,
\begin{eqnarray}
L_4^r = \frac{f_\pi^2}{m_\pi^2+2m_K^2} \left\{
 \frac{c}{4}(2\mu_\pi + \mu_K) - \frac{m_\pi^2}{f_\pi^2} L_5^r 
 \right\} .
\end{eqnarray}
One can take any renormalization scale $\mu$ at around $K^*$ meson
mass. We specifically choose the value of the particle data group
(PDG) \cite{Beringer:1900zz}, namely $\mu=895.47$MeV.
If $c(=1-M_V^2/(g^2f_\pi^2))$ is obtained, $L_4^r$ and $L_5^r$ can
be fixed.

From the imaginary part of the self energy for $K^\ast$ meson
in Eq.(\ref{del_ks}), the decay width of $K^\ast$ is 
given by,
\bea
\Gamma_{K^*}(M_{K^\ast}^2)=\frac{1}{16\pi M_{K^*}}
\frac{\nu_{K\pi}^3(M_{K^\ast}^2)}{M_{K^\ast}^4}
\left( \frac{M_V^2}{4gf_\pi^2} \right)^2 ,
\label{eq:GKstar}
\eea
where $\nu_{K \pi}$ is defined in Eq.(\ref{eq:nuPQ}). 
Once $M_V$ is determined, 
$g$ can be fixed with the decay width $K^\ast$ ($\Gamma_{K^*}$)
and $M_{K^*}$.
The relations among $Z^r_V, C_1^r, C_2^r$ are derived by 
the conditions for the pole masses of $K^*$ and $\rho$ mesons. 
We define $K^*$ and $\rho$ meson masses as the momentum 
squared ($Q^2$) for which the real parts of the inverse propagators
vanish,
\bea
M_V^2 + {\rm Re}[\delta A_{K^*}(Q^2=M_{K^*}^2;C_1^r,C_2^r)] &=& 0, \\
M_V^2 + {\rm Re}[\delta A_{\rho}(Q^2=M_{\rho}^2;C_1^r,C_2^r)] &=& 0.
\eea
Solving the above equations, one obtains $C^r_1$ and $C^r_2$,
\bea
C^r_1 &=& \frac1{\Delta_{K\pi}} \left\{ Z^r_V \Delta_{K^* \rho}
 - {\rm Re}[\Delta A_{K^*}(M_{K^*}^2)]
 + {\rm Re}[\Delta A_{\rho}(M_{\rho}^2)] \right\} , \\ 
C^r_2 &=& -\frac1{2m_K^2 + m_\pi^2 } \left\{ M_V^2 - Z^r_V M_{\rho}^2
 + {\rm Re}[\Delta A_{\rho}(M_{\rho}^2)]
 + C^r_1 m_{\pi}^2 \right\} ,
\eea
where,
\bea
\Delta A_{K^*} &=& -\frac34 \left( \frac{M_V^2}{gf_\pi^2} \right)^2
 \left[ Q^2(M_{K\pi}^r + M_{K\eta_8}^r) \right. 
 \left. -L_{K\pi}-L_{K\eta_8} + \frac{f_\pi^2}{2}(\mu_\pi + 2\mu_K 
 + \mu_{\eta_8}) \right], \\
\Delta A_{\rho} &=& - \left( \frac{M_V^2}{gf_\pi^2} \right)^2
 \left[ Q^2 \left( M_{\pi}^r + \frac12 M_{K}^r \right)
 + f_\pi^2 \left( \mu_\pi + \frac12 \mu_K \right) \right] .
\eea
From the condition for the residue of the vector meson propagator (\ref{prop}), 
$Z^r_V$ is written as follows,
\bea
Z^r_V = 1 + \left. \frac{d {\rm Re}[\Delta A_{K^*}(Q^2)]}{dQ^2}
\right|_{Q^2=M_{K^*}^2} .
\label{zrq}
\eea

We use $\rho$ meson mass of the PDG value \cite{Beringer:1900zz}.
For $K^*$ meson mass and the decay width, we fix them
with the hadronic mass spectrum of $\tau \to K \pi \nu$.
Instead of using
$g$ and $Z_V$ as the fitting parameters, one can use the decay width 
$\Gamma_{K^*}$ and the mass $M_{K^*}$ .
 
One can write $Z_V^r, C_3^r, C_4^r$ and $C_5^r$ in terms of 
$K_1^r, K_2^r, K_3^r$ and $L_9^r$ with Eqs. (\ref{zrcr}) and
(\ref{c5}). Since $K_3^r$ is related to $Z_V^r$ with Eq. (\ref{zrcr})
and $Z_V^r$ is fixed with Eq. (\ref{zrq}), $K_3^r$ is already
determined by $M_{K^*}$. We note that the form factor of Eq. (\ref{FKs_V}) 
depends on the combination $C_3^r-4C_4^r$, which is written as, 
\begin{equation}
C_3^r - 4C_4^r = \frac{M_V^2}{g f_\pi^2} \left(
 K_1^r + K_2^r -K_3^r \frac{M_V^2}{g^2f_\pi^2}
\right) .
\end{equation}
One also notes that $C_5^r$ is written in terms of $K_1^r+K_2^r,
K_3^r$ and $L_9^r$. Therefore we choose 
$\{\Gamma_{K^\ast}, M_V, M_{K^*}, K_1^r+K_2^r, L_9^r \}$
as fitting parameters in the following analysis.

\begin{figure}[!ht]
  \begin{center}
    \includegraphics[height=7cm,keepaspectratio]{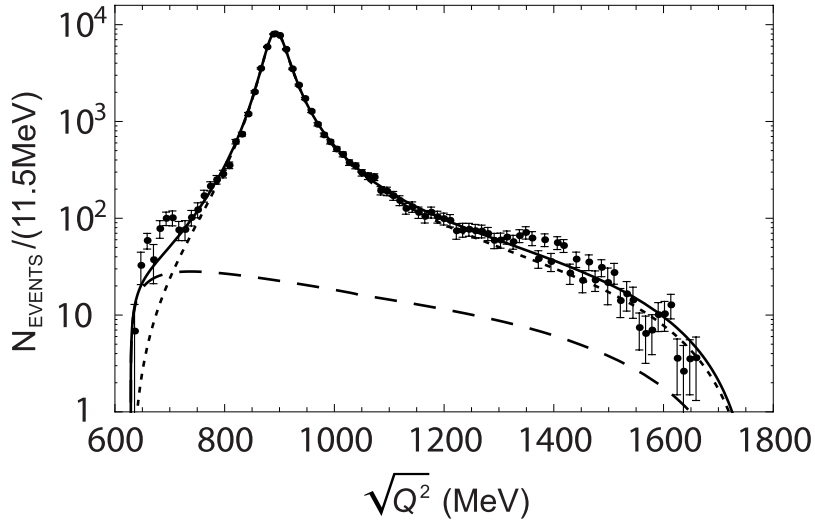}
  \end{center}
  \vspace{-0.5cm}
  \caption{The prediction of the decay distribution for $\tau^- \to K_s\pi^-\nu$.  The solid line corresponds to the prediction of our model.
The dotted line and the dashed line correspond to the distribution of the 
vector form factor and that of the scalar form factor, respectively.
The closed circles with the error bars are experimental 
   data  \cite{Belle:2007rf}.}
  \label{dir_Kpi}
\end{figure}

We fit them  by using the differential branching fraction of
the experimental data \cite{Belle:2007rf}.
The differential branching fraction for $KP\nu (P=\pi, \eta)$ is given by,
\bea
\frac{d {\rm Br}(\tau \to KP\nu)}{d\sqrt{Q^2}}
&=& \frac{1}{\Gamma_\tau} \frac{G_F^2 |V_{us}|^2}{2^5 \pi^3} 
   \frac{(m_{\tau}^2-Q^2)^2}{m_{\tau}^3} p_{K} \nn \\
&&\times \left[ \left( \frac{2 m_{\tau}^2}{3 Q^2}+\frac{4}{3} \right)  
 {p_{K}}^2  |F_V^{KP}(Q^2)|^2 
+\frac{m_{\tau}^2}{2} |F_S^{KP}(Q^2)|^2 \right] ,
\label{bran}
\eea
where $p_K$ is the momentum of $K$ in the hadronic center of mass
(CM) frame. The differential decay distribution for 
$\tau^- \to K_s\pi^-\nu$ is shown in Fig. \ref{dir_Kpi}. One can see the
peak of $K^*$ 
resonance around at $\sqrt{Q^2}\simeq 900$ MeV.  

The five parameters are determined by fitting the hadronic 
mass spectrum in the region 
$m_K+m_\pi \leq \sqrt{Q^2} \leq 1665$MeV with 90 bins data. 
We also use the PDG values \cite{Beringer:1900zz}, $m_{\pi^\pm}, 
f_{\pi^\pm}, m_{K^0}, m_\eta, m_{\eta'}, m_\tau$ as inputs. 
The set of parameters leading to the smallest $\chi^2/$n.d.f value
are fixed by
\begin{eqnarray}
 \Gamma_{K^\ast}&=&48.68{\rm MeV},\quad M_V=954.0{\rm MeV},\quad 
 M_{K^*}=895.4{\rm MeV} , \nonumber \\
 K_1^r+K_2^r&=&0.04517,\quad L_9^r=5.068\times 10^{-3} , 
\end{eqnarray}
where the obtained
$\chi^2/{\rm n.d.f.}$ is $152.3 /85$. 
The other parameters are shown in Table \ref{fit_para0}.
We also note $1-M_V^2/(2g^2f_\pi^2)=0.2688$ for this case. 
It implies that the relation of the vector meson dominance,
$M_V^2=2g^2 f_\pi^2$, is slightly violated.
\begin{table} [!ht]
 \caption{ Numerical values of the fitted parameters.}
 \label{fit_para0}
 \begin{ruledtabular}
 \begin{tabular}{|cc|cc|cc|}
  $g$ & 8.582 & $C_2^r$ & $-0.7772$ & $L_4^r$ & $2.265\times 10^{-4}$ \\ \hline 
  $Z_V^r$ & 0.8276 & $C_3^r-4C_4^r$ & 0.1811 
          & $L_5^r$ & $2.313\times 10^{-3} $\\ \hline
  $C_1^r$ & 0.2980 & $C_5^r$ & $-1.516\times 10^{-3}$  &  & 
 \end{tabular}
 \end{ruledtabular}
\end{table}

Table \ref{fit_para} shows the fitted values of $M_{K^*},\Gamma_{K^*}$
and the prediction of the branching fraction of our model. The corresponding 
experimental 
values are also shown.  We have not shown the error for the branching fraction, since the 
systematic error in each bin is not known.
We also show the obtained slope parameter for $K_{e3}$ decay defined in Eq.(\ref{lambda}).
$\lambda_+$ is the slope parameter given by the linear
expansion coefficient of the $K\pi$ vector 
form factor (\ref{F_V}), 
\begin{equation}
F_V^{K\pi}(Q^2) \simeq F_V^{K\pi}(0)
 \left( 1 + \lambda_+ \frac{Q^2}{m_\pi^2} \right),
\label{fv_kpi0}
\end{equation}
where,
\begin{eqnarray}
\lambda_+ = \frac{m_\pi^2}{F_V^{K\pi}(0)} \left. 
 \frac{dF_V^{K\pi}(Q^2)}{dQ^2} \right|_{Q^2=0} .
\label{lambda}
\end{eqnarray}

\begin{table} [!ht]
 \caption{Fitting parameters ($M_{K^\ast}, \Gamma_{K^\ast}$)
 and the corresponding predictions for the branching fraction and the slope parameter.
The bottom line denotes the experimental values.}
 \label{fit_para}
 \begin{ruledtabular}
 \begin{tabular}{|c|c|c|c|c|}
  & $M_{K^*}$(MeV)  & $\Gamma_{K^*}$(MeV) 
  & Br($\tau^- \to K_S\pi^-\nu_\tau$) & $\lambda_+$ \\
\hline 
  & 894.5 & 48.67 & 0.4023$\%$ & 0.02236 \\ \hline
Exp. & $895.47\pm0.20\pm0.74$
     & $46.2 \pm0.6\pm1.2$  
     & $(0.404\pm0.002\pm0.013)\%$ 
     & $0.02485\pm0.00163\pm0.00034$ 
 \end{tabular}
 \end{ruledtabular}
\end{table}

\begin{figure}[ht!]
\begin{flushleft}
 \begin{minipage}{75mm}
   \begin{center}
 \includegraphics[height=50mm,keepaspectratio]{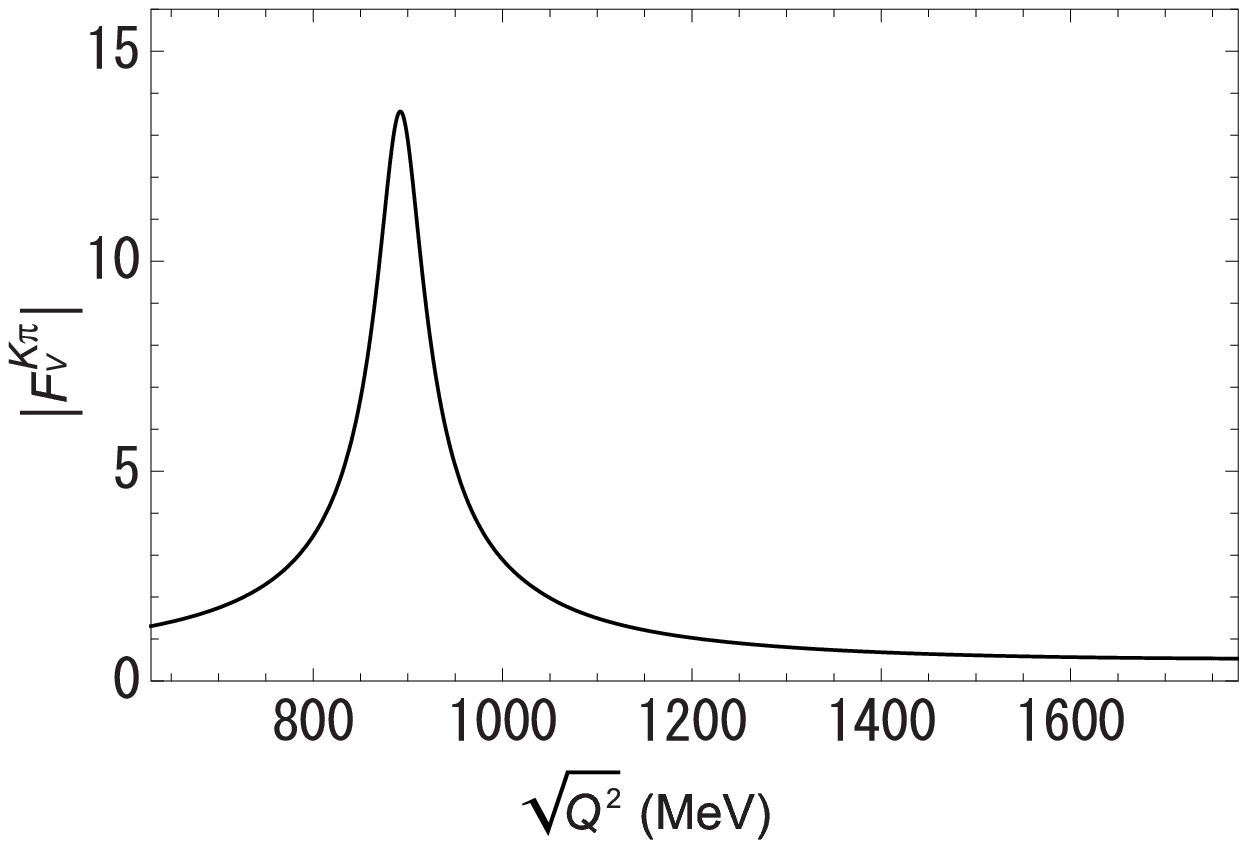}
 \caption{The absolute value of the vector form factor.}
 \label{abs_fv}
   \end{center}
 \end{minipage}
\hspace{10mm}
 \begin{minipage}{75mm}
 \begin{center}
\includegraphics[height=50mm,keepaspectratio]{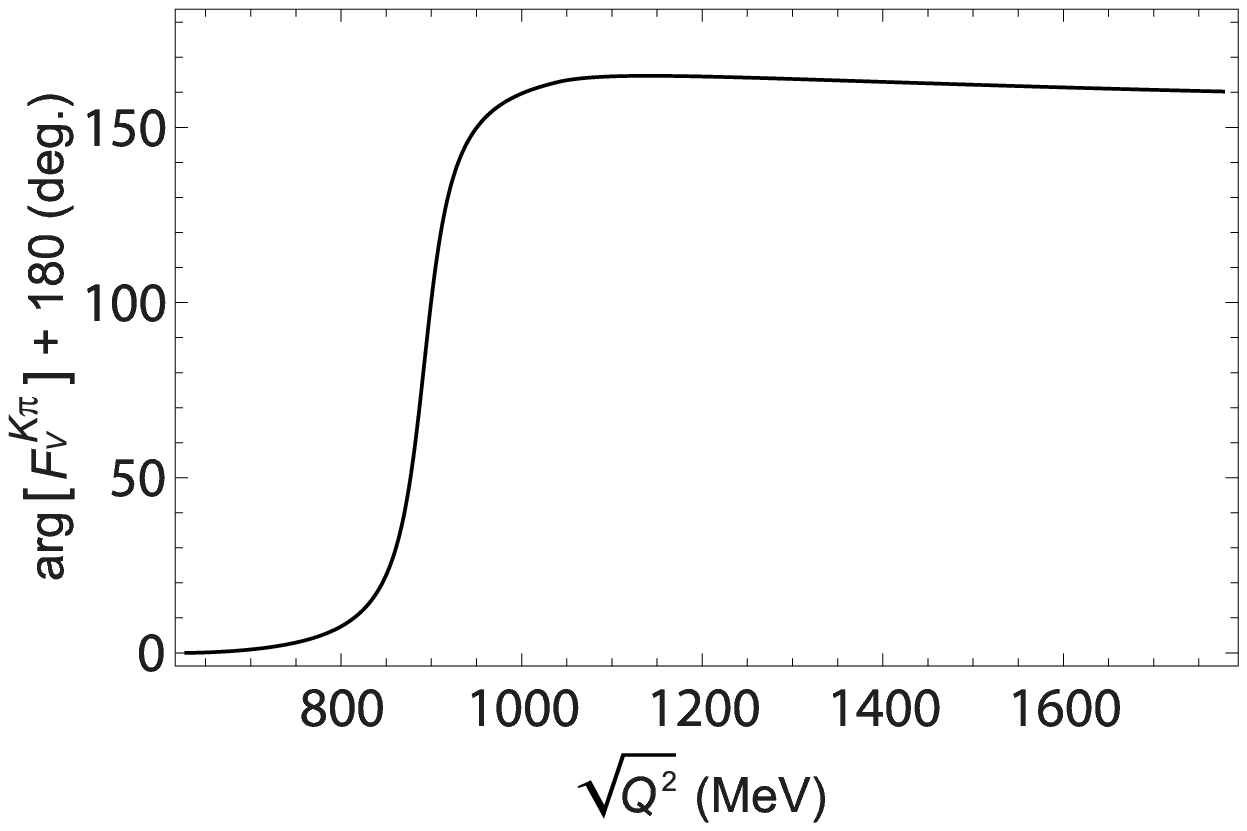}
\caption{The argument of the vector form factor. We added 180 deg. to ${\rm arg} F_V$.}
 \label{arg_fv}
 \end{center}
 \end{minipage}
\end{flushleft}
\end{figure}
\begin{figure}[ht!]
\begin{flushleft}
 \begin{minipage}{75mm}
 \begin{center}
 \includegraphics[height=50mm,keepaspectratio]{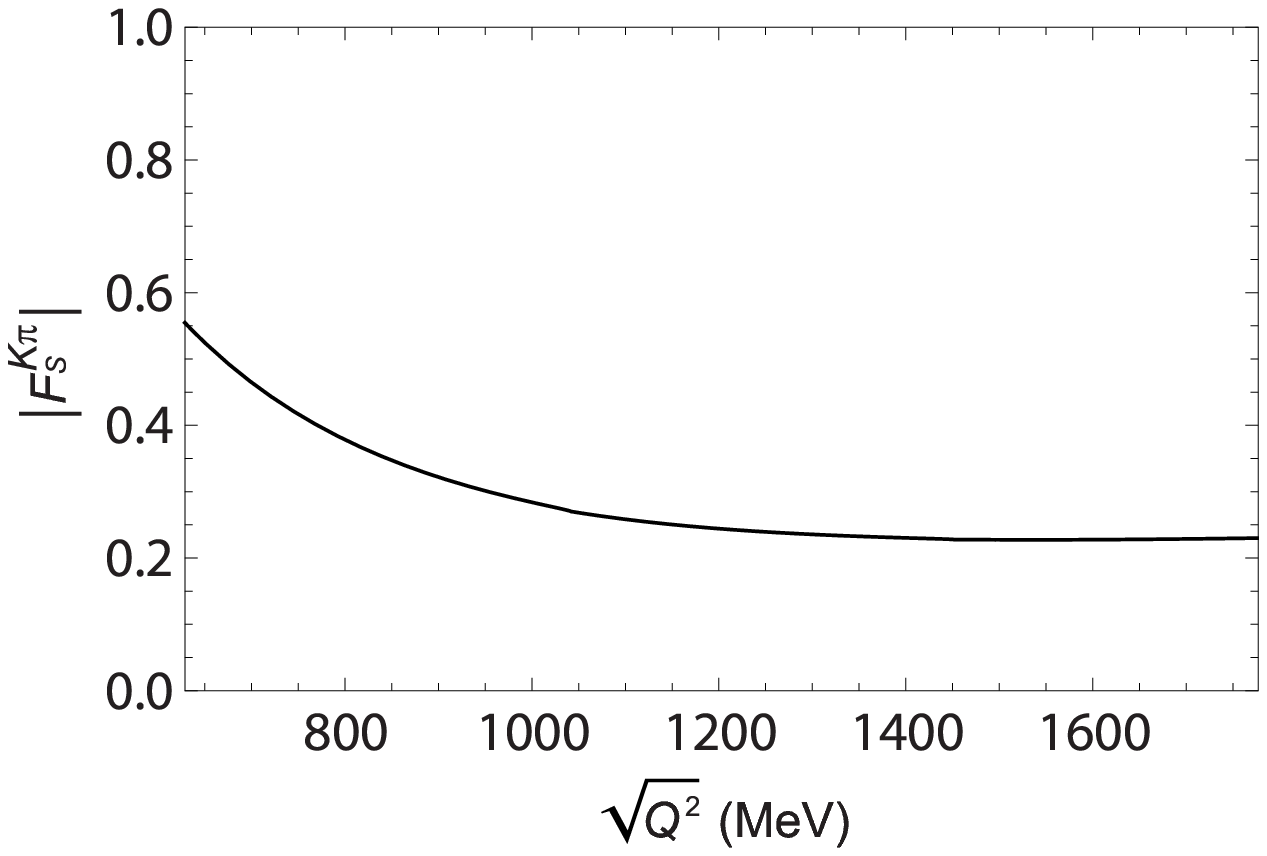}
 \caption{The absolute value of the scalar form factor.}
 \label{abs_fs}
 \end{center}
 \end{minipage}
 \hspace{10mm}
 \begin{minipage}{75mm}
 \begin{center}
  \includegraphics[height=50mm,keepaspectratio]{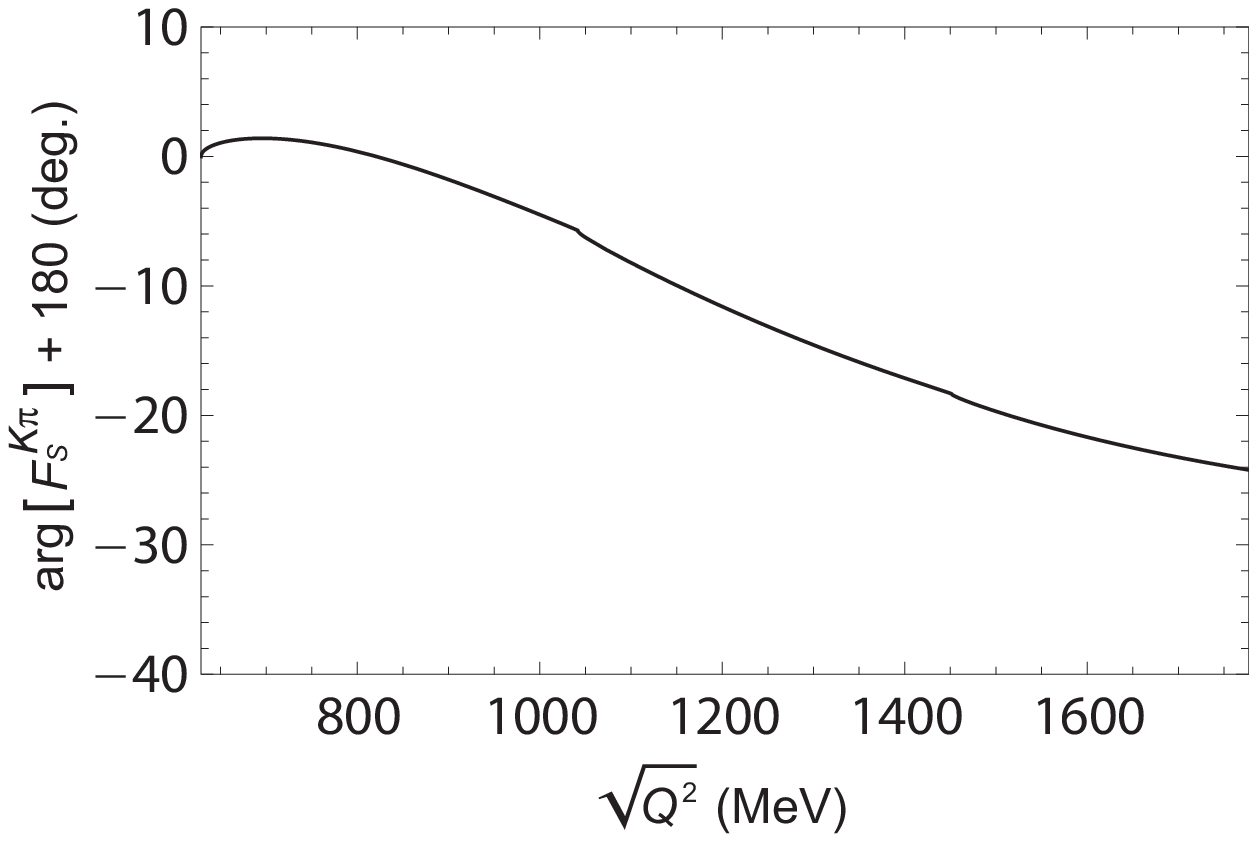}
 \caption{The argument of the scalar form factor.}
 \label{arg_fs}
 \end{center}
 \end{minipage}
\end{flushleft}
\end{figure}

   We investigate the property of the form factors obtained in the present work.
Using the fixed parameters, we show the absolute value and argument 
for the vector form factor in Figs. \ref{abs_fv} and 
\ref{arg_fv}.
In the absolute value of the vector form factor $|F_V^{K\pi}|$, the effect 
of $K^*$ resonance is dominant at $\sqrt{Q^2}\simeq M_{K^*}$. Furthermore, 
the effect of $K^*$ resonance is seen in 
the argument of vector form factor (arg$[F_V^{K\pi}]$), because it changes about 
$180^{\circ}$ near $\sqrt{Q^2} \simeq M_{K^*}$. This property
are also seen in $K\pi$ scattering \cite{SLAC-PUB-4260}.
  However, the behavior of arg$[F_V^{K\pi}]$ for large invariant mass region,
$\sqrt{Q^2} \gtrsim 1200$MeV is different from the one in \cite{SLAC-PUB-4260}, 
since our model does not include
the higer resonances, $K^*(1410)$ and $K^*(1790)$.
Figures \ref{abs_fs} and \ref{arg_fs} show the 
absolute value and argument for the scalar form factor, respectively. 
  The absolute value of the scalar form factor is smaller than the
absolute value for the vector form factor, since there is no $K^*$ pole
in $F_S^{K\pi}$ as shown in Eq. (\ref{FKs_S}). As increasing the invariant mass,
the argument for the scalar form factor decreases.

We study $\tau^- \to K^-\eta \nu$ decay using the parameters
fixed with $\tau^- \to K_s \pi^- \nu$ decay. The form factors for $K\eta$ 
are given in Appendix E. Figure \ref{dir_Keta} shows the
prediction of the decay distribution for $\tau^- \to K^-\eta\nu$.
It is found that the contribution of vector form factor is dominant. 
The predicted branching fraction 
for $\tau^- \to K^-\eta \nu$ decay is $2.114\times 10^{-4}$.
Since the experimental results are Br$(\tau^- \to K^-\eta \nu)=
(1.52 \pm 0.08)\times 10^{-4}$ \cite{Beringer:1900zz}, our prediction 
is larger than the experimental data. 
   We note that the predicted branching fractions 
for $\tau^- \to K_S \pi^- \nu$ and $\tau^- \to K^-\eta \nu$ decays are 
$4.030\times 10^{-3}$ and $1.157\times 10^{-4}$ respectively with the other
parameter set of parameters which is obtained by 67 bins data fitting 
($m_K+m_\pi \leq \sqrt{Q^2} \leq 1400.5$MeV).

\begin{figure}[!ht]
  \begin{center}
    \includegraphics[height=7cm,keepaspectratio]{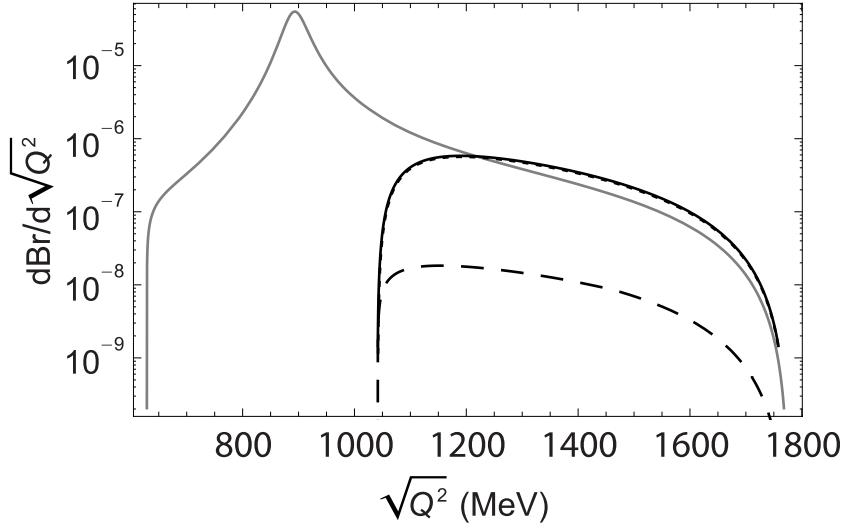}
  \end{center}
  \vspace{-0.5cm}
  \caption{The hadronic invariant mass 
distribution for $\tau^- \to K^-\eta\nu$ decay.
The solid line and the gray solid line correspond to
the hadronic invariant mass distribution for 
   $\tau^- \to K \eta \nu$ decay and that for $\tau \to K_s \pi \nu$
 decay, respectively.
   The dotted line and the dashed line are 
the vector form factor contribution and the scalar form factor 
   contribution of $\tau \to K \eta \nu$ decay, respectively.}
  \label{dir_Keta}
\end{figure}

\begin{figure}[htbp]
\begin{center}
\includegraphics[width=9cm]{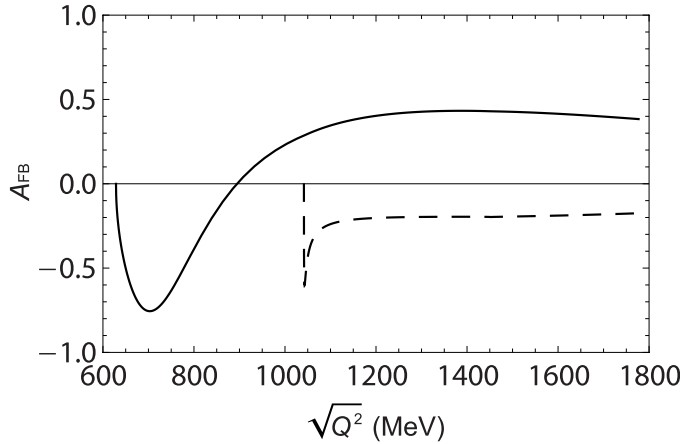}
\end{center}
\caption{The predictions 
of the forward-backward asymmetries of $\tau \to
K \pi \nu $ and
$\tau \to K \eta \nu $ decays.
The solid line and dashed line correspond to the forward-backward asymmetry of $\tau \to K \pi \nu$ decay and that of the $\tau \to K \eta \nu$ decay, respectively.}
\label{FBASM}
\end{figure}

We also consider the forward-backward
asymmetry \cite{Beldjoudi:1994hi} for $\tau \to K P \nu $ decay.
The double differential rate of the unpolarized
$\tau$ decay \cite{Kuhn:1996dv} is given by
\bea
\frac{d {\rm Br}}{d \sqrt{Q^2} d \cos \theta}
&=& \frac{1}{\Gamma}
\frac{G_F^2|V_{us}|^2}{2^5 \pi^3} \frac{(m_{\tau}^2-Q^2)^2 p_K}{m_{\tau}^3}
\Biggr\{ \left(\frac{m_\tau^2}{Q^2}\cos^2 \theta
+\sin^2 \theta \right) p_K^2|F_V^{KP}(Q^2)|^2  \nn \\
&&  +\frac{m_{\tau}^2}{4} |F_S^{KP}|^2
- \frac{m_{\tau}^2}{\sqrt{Q^2}} p_K \cos \theta
{\rm Re}[F_V^{KP}(Q^2) F_S^{KP}(Q^2)^{\ast}] \Biggr\},
\eea
where $\theta$ is the scattering angle of kaon with respect to
the incoming $\tau$ in the hadronic CM frame.
The forward-backward asymmetry extracts the interference term
of the vector form factor and the scalar form factor.
\bea
A_{\rm FB}(Q^2)&=&\frac{\int^1_0 d \cos \theta \frac{d {\rm Br}}{d \sqrt{Q^2}
d \cos \theta}-\int^0_{-1} d \cos \theta 
\frac{d {\rm Br}}{d \sqrt{Q^2}d \cos \theta}}
{\frac{d {\rm Br}}{d \sqrt{Q^2}}} \nn \\
&=&-\frac{ \frac{p_K}{\sqrt{Q^2}}
\frac{|F_S^{KP}|}{|F_V^{KP}|} \cos \delta^{KP}_{\rm st} 
}{
\left(\frac{2 m_{\tau}^2}{3 s}+\frac{4}{3}\right) \frac{p_K^2}{m_{\tau}^2}
+\frac{1}{2} |\frac{F_S^{KP}}{F_V^{KP}}|^2 },
\label{eq:FB}
\eea
with $\delta^{KP}_{\rm st}={\rm arg}.(\frac{F_V^{KP}}{F_S^{KP}}).$
As we can see from Eq.(\ref{eq:FB}), the forward-backward 
asymmetry is determined by the ratio of the
scalar and the vector form factors. It is also proportional 
to cosine of the strong phase shift $\delta_{\rm st}^{KP}$.
The forward-backward asymmetries
for $K \pi$ and $K \eta$ cases are shown in 
Fig. \ref{FBASM}.
As can be seen in Fig. \ref{FBASM}, the forward-backward asymmetry for 
$K \pi$ case
is large below $K^*$ resonance and reaches to $70\%$.
Here the decay distribution for $\tau^- \to K_s \pi^- \nu$
is identical to that of $\tau^- \to K^- \pi^0 \nu$ by taking the limit for 
$\epsilon_K$ of $K^0 \overline{K^0}$ mixing zero. In Fig. \ref{FBASM},
we have evaluated the forward-backward asymmetry for
$\tau^- \to K^- \pi^0 \nu$ as that for $K\pi$ case.

\section{Two Higgs doublet model with  CP violation}
As an example of new physics beyond the SM, 
we investigate a two Higgs doublet model with explicit CP violation. 
(See for example, Ref. \cite{Branco:2011iw} for a recent review of two 
Higgs doublet model.) 
The model is classified as type II two Higgs doublet model. 
$Z_2$ parity is assigned so that only a Higgs doublet $\Phi_2$ is coupled to up type quarks and another Higgs doublet $\Phi_1$ is coupled with
down type quarks. 
For the charged leptons, they have Yukawa couplings with the same Higgs doublet which the down type quarks interact with.
$Z_2$ symmetry is softly broken in Higgs sector. By taking the
soft breaking mass squared parameter small, one can naturally obtain the large ratio of vacuum 
expectation values (VEVs) of two Higgs doublets
. The idea of Ref.
\cite{Hashimoto:2004kz} is that this large ratio of the Higgs VEVs is the origin of the isospin breaking of the third generation of the quarks. In such model, the Higgs with small VEV has
enhanced Yukawa couplings to down type quarks and charged leptons. Therefore,
$\tau$ lepton and bottom quark can be good probes investigating the extra Higgs
doublet with the small VEV. 

The well known effect of CP violation of the two Higgs doublet model is 
CP even and CP odd Higgs mixing 
\cite{Pilaftsis:1997dr, Pilaftsis:1999qt}. 
In the large limit of the ratio of 
Higgs VEVs,
among three neutral Higgs, the SM like CP even Higgs is decoupled 
from  the other two Higgs bosons. Therefore, in good approximation,  
CP even and CP odd Higgs mixing occurs among two Higgs bosons
in the sector of the Higgs with the small VEV. 
We investigate how the CP violating mixing of
the neutral Higgs sector leads to some observable effect on charged Higgs 
Yukawa coupling. We also explicitly show
how it generates the direct CP violation of $\tau$ decays.
 For this purpose, we compute one loop corrections to masses of 
the charged leptons and down type quarks. One finds the one loop corrected
mass is flavor diagonal and a small CP violating chiral phase due to
the CP even and CP odd Higgs mixing is generated.
To remove the phase of one loop corrected mass, one needs to carry out 
the chiral rotation. After the chiral rotation, 
CP violating phase in charged Higgs sector arises.
The phase is due to the CP violation of Higgs sector 
which is the different origin from Kobayashi Maskawa phase 
\cite{Kobayashi:1973fv}.

After all, the relative CP violating phase difference between the charged 
current interaction of $W$ boson and charged Higgs interaction arises as,
\bea
{\cal L}\sim \overline{\nu_L} \gamma_\mu \tau_L W^{\mu +}+ \overline{\nu_L}
\tau_R H^+ e^{-2 i \phi_\tau}.
\eea
The phase $\phi_\tau$ vanishes if CP even and CP odd Higgs mixing 
angle $\theta_{AH}$ vanishes.
The phase $\phi_\tau$ can be measured by direct CP violation of
$\tau^{\pm}$ decays. The decays 
go through the intermediate states $W^-$ and $H^-$
which are converted to a common hadronic final state $(K, \pi)$.
Schematically, the process goes as, 
\bea
\tau \rightarrow \Biggl{\{}
\begin{array}{c}
\nu_L + W^{- \ast} \\ 
\nu_L + H^{- \ast}
\end{array}  \Biggr{\}} \rightarrow K^- \pi^0 + \nu.
\eea
To measure the phase $\phi_\tau$, the angular analysis of 
the decay distributions of $\tau \to K \pi \nu$ is useful.  
The direct CP violation arises in the interference of two amplitudes with 
both weak phase difference and strong phase difference.
In the $\tau \to K \pi \nu$ decays, 
the interference of two amplitudes with different angular momentum of 
$K^- \pi^0$ , i.e., $l=1$ and $l=0$ can take place. 
The difference of the angular distribution of $\tau^- \to K^- \pi^0 \nu$ and 
its CP conjugate $\tau^+ \to K^+\pi^0 \bar{\nu} $ is sensitive to the CP 
violating phase described above.
As we have shown in \cite{Kimura:2008gh}, 
the forward-backward CP asymmetry is a good observable
for the CP violation.

The Higgs potential of two Higgs doublet model with softly broken $Z_2$
symmetry is given as,
\bea
V_{\rm tree}&=&\sum_{i=1,2}\left( m_{i}^2 \Phi_i^\dagger \Phi_i+
\frac{\lambda_i}{2} (\Phi_i^\dagger \Phi_i)^2 \right)
-m_{3}^2 (\Phi_1^\dagger \Phi_2 + h.c.)+\lambda_3  (\Phi_1^\dagger \Phi_1)(\Phi_2^\dagger \Phi_2) \nn \\
&&+\lambda_4 |\Phi_1^\dagger \Phi_2|^2
+ \frac{1}{2} \lambda_5 \Bigl{[}e^{i \theta_5}(\Phi_2^\dagger \Phi_1)^2 + e^{-i \theta_5}(\Phi_1^\dagger \Phi_2)^2 \Bigr{]},
\label{eq:tree}
\eea
where under $Z_2$ transformation, the Higgs fields transform as,
\bea
\Phi_1 \rightarrow -\Phi_1, \quad  \Phi_2 \rightarrow \Phi_2.
\eea
$\theta_5$ is a CP violation parameter of Higgs sector.
One may write the vacuum expectation values with three
order parameters \cite{Morozumi:2011zu},
\bea
\langle \Phi_1 \rangle =\frac{v}{\sqrt{2}}\begin{pmatrix}
0 \\
\cos \beta
\end{pmatrix}, \quad 
\langle \Phi_2 \rangle =\frac{v}{\sqrt{2}} \begin{pmatrix} 
0 \\
\sin \beta  \end{pmatrix} e^{-i \theta^\prime}.
\eea
The three order parameters are determined by the stationary conditions.
For large $\tan \beta$, the solution can be written approximately as,
\bea
v^2  &\simeq&  -\frac{2 m_2^2}{\lambda_2},\nn \\
\cos \beta
&\simeq&  \frac{m_3^2}{\left\{m_1^2+\frac{v^2}{2}(\lambda_3+\lambda_4) \right\}
\cos \theta^\prime+ \frac{v^2}{2} \lambda_5 
\cos(\theta_5+\theta^\prime)}, \nn \\
\frac{\sin (\theta_5+ \theta^\prime)}{\sin \theta^\prime}
&\simeq&\frac{\lambda_3+\lambda_4-\frac{m_1^2}{m_2^2} \lambda_2}{\lambda_5},
\label{eq:relation}
\eea
where only the leading terms 
with respect to the expansion of the soft breaking 
parameter $\frac{m_3^2}{m_{1}^2}$ are shown.
When  $\theta_5$ is not vanishing, the neutral Higgs bosons with
definite CP parities,i.e.,
CP even ($H$) and CP odd Higgses ($A$) are not mass eigenstates. 
Their mixing angle is sensitive to the CP violation of the Higgs
sector.
In large $\tan \beta$
limit, the mass matrix of the three neutral Higgs becomes,
\bea
{\cal L}_{\rm mass}=-\frac{v^2}{4}(h,H,A) \begin{pmatrix} a_{11} & 0 & 0 \\
0 & a_{22} & a_{23} \\
0 & a_{23} & a_{33} \end{pmatrix} 
\begin{pmatrix} h \\ H \\ A \end{pmatrix},
\eea
where $a_{12}
$ and $a_{13}$ are subleading of the expansion of 
$\cos \beta$ and can be neglected in large $\tan \beta$ limit.
Therefore in the limit, one can simply diagonalize $2 \times 2$ matrix.
For the purpose, one introduces the mixing angle $\theta_{AH}$,
\bea
H&=&\cos \theta_{AH} H_3 + \sin \theta_{AH} H_2, \nn \\
A&=&\cos \theta_{AH} H_2 -\sin \theta_{AH} H_3,
\eea
where $H_2$ and $H_3$ are mass eigen states.
The other matrix elements in small $\cos \beta$ limit are,
\bea
a_{11} & \simeq & 2 \lambda_2,\nn \\
a_{33} & \simeq &\left\{ \frac{\sin(\theta_5+\theta^\prime)}{\sin \theta^\prime}-\cos (\theta_5+2 \theta^\prime) \right\} \lambda_5, \nn \\
a_{22} &\simeq & \left\{ \cos \theta_5+ \cos 2\theta^\prime\frac{\sin (\theta^\prime+\theta_5)}{\sin \theta^\prime} \right\} \lambda_5, \nn \\
a_{23} &\simeq & -\lambda_5 \sin (\theta_5+2 \theta^\prime). 
\eea
Then one finds the mixing angle is given by,
\bea
\theta_{AH}= \frac{\theta_5}{2}+\theta^\prime.
\eea
In the same limit, the masses of 
all the Higgs bosons are;
\bea 
{\cal L}_{\rm mass}= -\frac{M_{h}^2}{2}  h^2- \frac{M_{H_2}^2}{2} H_2^2
-\frac{M_{H_3}^2}{2} H_3^2-M_{H^+}^2 H^+ H^-,
\eea
with,
\bea
M_{h}^2&=&\lambda_2 v^2, \nn \\
M_{H^+}^2&=& m_1^2+\frac{\lambda_3}{2} v^2, \nn \\
M_{H_2}^2&=& m_1^2+\frac{v^2}{2}(\lambda_3+\lambda_4-\lambda_5), \nn \\
M_{H_3}^2&=&m_1^2+\frac{v^2}{2}(\lambda_3+\lambda_4+\lambda_5).
\label{eq:massf}
\eea
\begin{figure}[bhtp]
\begin{center}
\includegraphics[height=6cm]{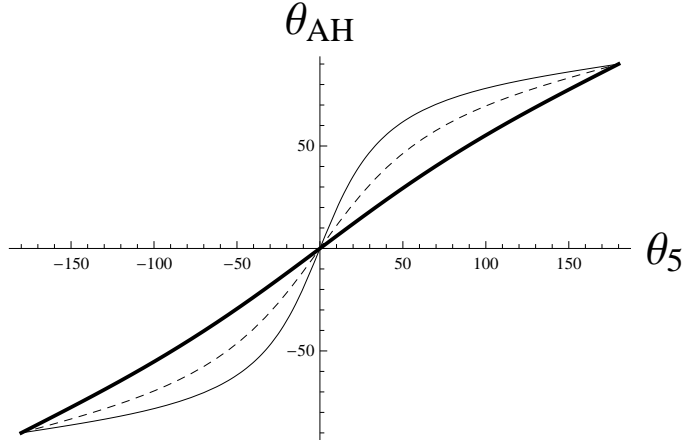}
\caption{ The CP even and odd
Higgs mixing angle $\theta_{AH}$ as a function of 
CP violating parameter $\theta_5$. The thin solid line, the dashed line, and
the thick solid line correspond to 
$\frac{M_{H_3}}{M_{H_2}}=2$, $1.5$, and
$1.1$, respectively.}
\end{center}
\label{tAH}
\end{figure}

Using the relations in Eq.(\ref{eq:relation}) and the mass formulae of
Higgs bosons in Eq.(\ref{eq:massf}), one can write the formulae
$\cos \beta$ and $\theta_{AH}$ as follows;
\bea
\cos \beta&=& \frac{m_3^2}{\sqrt{M_{H_3}^4 \cos^2 \theta_{AH} + 
M_{H_2}^4 \sin^2 \theta_{AH}}}, \label{eq:cb} \\
\theta_{AH}&=&\arctan\left(
\frac{M_{H_3}^2}{M_{H_2}^2}\tan \frac{\theta_5}{2}\right). \label{eq:tAH}
\eea
In Fig.12, we have shown the mixing angle $\theta_{AH}$
as a function of CP violating parameter $\theta_5$ of the Higgs potential
as given by Eq.(\ref{eq:tAH}).
One can see when the mass splitting of $H_2$ and $H_3$ are large, $\theta_{AH}$
tends to deviate from the line of $\theta_{AH}=\frac{\theta_5}{2}$, 
which leads to $\theta^\prime$ is non-vanishing. 
\begin{figure}[htbp]
\begin{center}
\includegraphics[height=3cm]{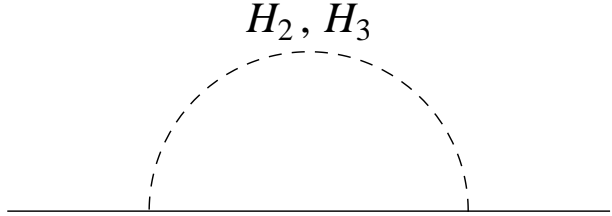}
\caption{One loop corrections to the self energies of down type quarks and 
charged leptons due to neutral Higgs exchanges.}
\end{center}
\label{oneloopmass}
\end{figure}
Next we compute the one loop corrected mass due to $H_2$ and $H_3$.
Yukawa couplings of them to down type quarks and charged leptons
can be written as,
\bea
{\cal L}_Y&=&\frac{H_2}{v}\left[ \tan \beta (\overline{e_i} i\gamma_5 e^{-i \gamma_5 \theta_{AH}} m_{li} e_i
+\overline{d_i} i \gamma_5 e^{-i \gamma_5 \theta_{AH}} m_{di} d_i )
                                                \right] \nn \\
           &&+\frac{H_3}{v}\left[ \tan \beta (\overline{e_i} e^{-i \gamma_5 \theta_{AH}} m_{li} e_i
           +\overline{d_i} e^{-i \gamma_5 \theta_{AH}} m_{di} d_i) \right].
\eea
Note that the Yukawa couplings of $H_2$ and $H_3$ have an enhancement 
factor $\tan \beta$.  
The CP violation of the Yukawa couplings are written in terms of the chiral
phase, $e^{-i\gamma_5\theta_{AH}}$.  
One defines the one loop corrected masses for down type quarks and charged leptons as,
\bea
{\cal L}_{mass}=-\overline{l_{Li}} M_{li} l_{Ri}
-\overline{d_{Li}} M_{d_i} d_{Ri}+ h.c..
\eea
The corrections are evaluated by computing Feynman diagrams  Fig.13
and the result is,
\bea
\Sigma_i|_{1 loop}=&& \left(\frac{m_i \tan \beta}{v}\right)^2
\Bigl{[}\frac{\slashed{p}}{16 \pi^2} (\frac{1}{\epsilon}-\gamma+ \ln 4 \pi)
-\frac{m_i}{16 \pi^2}(\log \frac{M_{H_2}^2}{M_{H_3}^2}) e^{-2 i\theta_{A} \gamma_5}
\Bigr{]} \nn \\
&&-(Z_i-1) \slashed{p} +(Z_{m_i}Z_i-1) m_i,
\eea
where $i$ denote the charged lepton or doun type quark.
We have ignored the finite contribution suppressed by a factor of $\frac{m_i^2}{M_{H_2}^2}$ and $\frac{m_i^2}{M_{H_3}^3}$.
In MSbar scheme, the counter terms are determined
as,
\bea
Z_i-1=-(Z_{m_i}-1)=\frac{1}{16 \pi^2} \left(\frac{m_i \tan \beta}{v}
\right)^2
(\frac{1}{\epsilon}-\gamma+ \log 4 \pi).
\eea
Therefore the one-loop corrected masses are finite and are given by,
\bea
M_{l_i}=m_{l_i}\left\{1-\left(\frac{m_{l_i}\tan \beta}{4 \pi v}\right)^2 \ln \frac{M_{H_2}^2}{M_{H_3}^2}
 e^{-2i \theta_{AH}}\right\}, \nn \\
M_{d_i}=m_{d_i}\left\{1-\left(\frac{m_{d_i}\tan \beta}{4 \pi v}\right)^2 \ln \frac{M_{H_2}^2}{M_{H_3}^2}
 e^{-2i \theta_{AH}}\right\}.
\eea
In order to remove the phases of the one loop corrected mass,
one need to perform the flavor diagonal chiral rotation,
\bea
&&l_{Ri} \rightarrow l_{Ri} e^{-i \phi_{l_i}},\quad l_{Li} \rightarrow l_{Li} 
e^{i\phi_{l_i}}, \nn \\
&&d_{Ri} \rightarrow d_{Ri} e^{-i \phi_{d_i}},\quad d_{Li} \rightarrow d_{Li} 
e^{i\phi_{d_i}},
\eea
where the phases $\phi_{l_i}$ and $\phi_{d_i}$ are given by,
\bea
\tan 2 \phi_{l_i}&=&\frac{\sin2 \theta_{AH} \left(\frac{m_{l_i}\tan \beta}{4 \pi v}\right)^2 \ln \frac{M_{H_2}^2}{M_{H_3}^2} }{1-
\left(\frac{m_{l_i}\tan \beta}{4 \pi v}\right)^2 \cos 2 \theta_{AH} 
\ln  \frac{M_{H_2}^2}{M_{H_3}^2}},
\label{eq:phitau}
\\
\tan 2 \phi_{d_i}&=&\frac{\sin2 \theta_{AH} \left(\frac{m_{d_i}\tan \beta}{4 \pi v}\right)^2 \ln \frac{M_{H_2}^2}{M_{H_3}^2} }{1-
\left(\frac{m_{d_i}\tan \beta}{4 \pi v}\right)^2 \cos 2 \theta_{AH} 
\ln  \frac{M_{H_2}^2}{M_{H_3}^2}}.
\label{eq:phib}
\eea
\begin{figure}[htbp]
\begin{center}
\includegraphics[width=9cm]{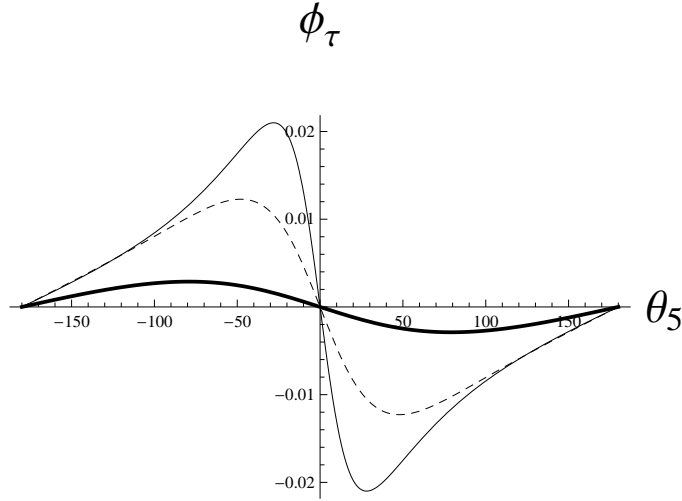}
\caption{
$\phi_{\tau}$ as a function of $\theta_5$. The unit of the angles
is in degree. The thin solid line,
the dashed line, and the thick solid line correspond to 
$\frac{M_{H_3}}{M_{H_2}}=2$, $1.5$, and $1.1$, respectively.}
\end{center}
\label{thtau152}
\end{figure}
In Fig.14, we have shown $\phi_{\tau}$ for different ratios 
of the Higgs mass $\frac{M_{H_3}}{M_{H_2}}$. 
The larger ratio leads to the larger value of $\phi_{\tau}$.

Now we study the effects of the CP violation of the Higgs mixing 
on $\tau$ lepton decays. The effective four Fermi interactions
from the SM contribution and from the charged Higgs
exchange are given by,
\bea
{\cal L}_{cc}&=&2 \sqrt{2} G_F V_{ji}^\ast
\Bigl{[}-\overline{d_{i L}} \gamma_\mu u_{j L}
\overline{\nu_L} \gamma^\mu \tau_L \nn \\
&&+\frac{m_\tau m_{di}\tan^2 \beta}{M_{H^+}^2}
\overline{d_{i R}}u_{j L}
e^{-2i (\phi_\tau-\phi_{di})}\overline{\nu_L} \tau_R \Bigr{]},
\label{eq:fourfermi}
\eea
where the relative phase $\phi_{\tau}-\phi_{d_i}$
of charged current interaction due to $W^-$ exchanged 
and charged Higgs interaction
$H^-$ arises.

 The forward-backward CP asymmetry in the two Higgs doublet model
can be obtained by replacing the SM scalar form
factor with the one including the charged Higgs contribution 
in Eq.(\ref{eq:FB}),
\bea
F_{S New}^{KP}\equiv
\left\{1-e^{-2i(\phi_\tau-\phi_s)}
\frac{Q^2 \tan^2 \beta}{M_{H^+}^2}\right\}F_S^{KP}.
\eea
By comparing the forward-backward asymmetry of $\tau^-$ and $\tau^+$,
one obtains the direct CP violation 
\cite{Kimura:2008gh},
\bea 
&& A_{FB}(\tau^- \to K^- P\, \nu)-\overline{A}_{FB}(\tau^+ \to K^+ P\, 
\overline{\nu})\nn \\
&&= -2 \sin \delta_{st}^{KP} \sin\{2 (\phi_\tau-\phi_s)\} 
\frac{Q^2 \tan^2 \beta}{M_{H^+}^2}
\frac{\frac{p_K}{\sqrt{Q^2}}\frac{|F_{S}^{KP}|}{|F_V^{KP}|}}
{(\frac{2 m_{\tau}^2}{3 Q^2}+\frac{4}{3}) \frac{p_K^2}{m_\tau^2}
+\frac{1}{2}\frac{|F^{KP}_{S New}|^2}{|F_V^{KP}|^2}},
\label{FBACPE}
\eea
where $P=\pi^0, \eta$.

\begin{figure}[htbp]
\begin{center}
\includegraphics[width=6cm]{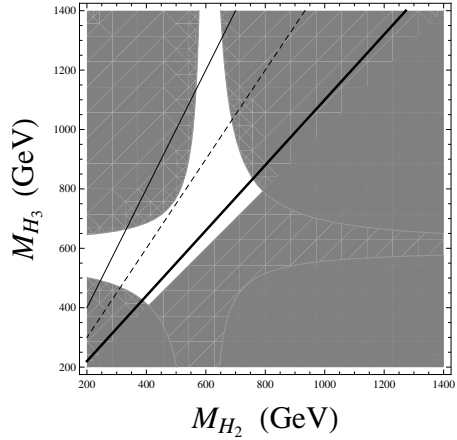}
\caption{The constraints on $T$ parameter in $(M_{H_2}, M_{H_3})$ plane.
The gray shaded regions are excluded. 
We choose $M_{H^+}=600$GeV. The upper bound on $T_{\rm New}$ 
is $0.15$ and the lower bound is $-0.068$, respectively. 
The thick solid line, the dashed line, and the thin solid line correspond to
$\frac{M_{H_3}}{M_{H_2}}=1.1, 1.5$, and $2$, respectively.}
\end{center}
\end{figure}
To predict the CP violation, we take account of the constraints
on the mass of the charged Higgs, $\tan \beta$ and the ratio of 
neutral Higgs masses $\frac{M_{H_3}}{M_{H_2}}$.
The lower limit of the charged Higgs mass is given as $M_{H^+} > 295$GeV 
obtained  from $B \to X_s \gamma$ \cite{Hewett:1992is,Barger:1992dy,Misiak:2006zs,
Bona:2009cj}.
Using  $B \to \tau \nu$ \cite{Hou:1992sy} and $B \to D \tau \nu$ 
\cite{Tanaka:1994ay}, the lower limit of the 
charged Higgs mass is constrained as $M_{H^+}>500$GeV for $\tan \beta \simeq
40$ \cite{Bona:2009cj}.
The ratio $\frac{M_{H_3}}{M_{H_2}}$ of the neutral Higgs masses can be
constrained from $T$ parameter.
$T$ parameter of the present model is computed as \cite{ElKaffas:2006nt},
\bea
T_{\rm New}=\frac{1}{16 \pi M_W^2 s_W^2}
\Bigl{[}F(M_{H^+},M_{H_3})+F(M_{H^+},M_{H_2})-F(M_{H_3},M_{H_2}) \Bigr{]},
\label{t_new}
\eea
where $F(m_a,m_b)$ is given by,
\bea
F(m_a,m_b)=\frac{m_a^2+m_b^2}{2}- \frac{m_a^2 m_b^2 \log \frac{m_a^2}{m_b^2}}{m_a^2-m_b^2}.
\eea 
In Eq.(\ref{t_new}), we take the limit; $\beta \to \frac{\pi}{2}$.
From Eq.(10.61) of Ref. \cite{Nakamura:2010zzi}, $T_{\rm New}=0.03\pm0.11$
for the SM Higgs boson mass $M_h=117$GeV case. We shift the SM reference 
point for the Higgs mass to $M_h=126$GeV \cite{:2012gk}, which amounts to 
the shift of $T_{\rm New}$ is $\frac3{8\pi c_W^2}\log\frac{126}{117}\simeq0.01$.
Therefore we adopt the following value for $T_{\rm New}$, 
\bea
T_{\rm New}=0.04 \pm 0.11.
\eea
With $M_{H^+}=600$GeV, the constraints on $(M_{H_2},M_{H_3})$ plane 
are shown in Fig.~15.

In Fig.~16, the forward-backward CP asymmetry in Eq.(\ref{FBACPE}) is shown.We neglect $\phi_s$ in the numerical calculation. We choose the charged Higgs
mass $M_{H^+}=600$GeV and $\tan \beta=40$, and $\frac{M_{H_3}}{M_{H_2}}=1.1,1.5$,and $2$ which satisfy the constraints studied. The CP asymmetry is as small
as $10^{-6} \sim 10^{-7}$.
Comparing the present result with the one with the two Higgs doublet model 
without natural flavor conservation \cite{Kimura:2008gh}, 
the asymmetry is much smaller
in the present model because it is the one-loop effect. 
\begin{figure}[htbp]
\begin{center}
\includegraphics[width=12cm]{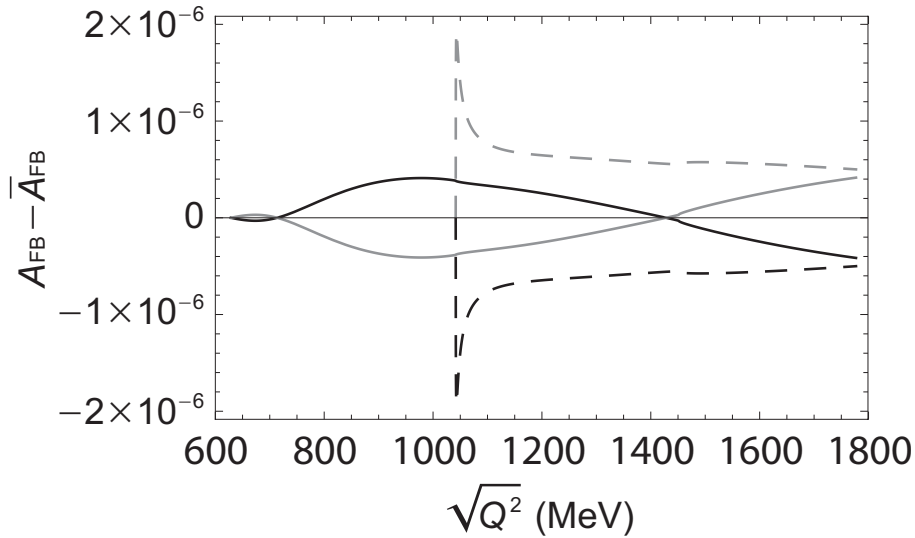}
\caption{
\small{Forward-backward CP asymmetry $A_{FB}-\bar{A}_{FB}$
as a function of hadronic invariant mass $\sqrt{Q^2}$.
The solid lines correspond to the ones of $\tau \to K \pi \nu $
decay
and the long dashed lines correspond to the ones of 
$\tau \to K \eta \nu$ decay.
The gray lines 
correspond to $\phi_\tau=0.02^{\circ}$ and the black
lines correspond to $\phi_{\tau}=-0.02^{\circ}$. We choose the ratio
of the neutral Higgs mass as 
$\frac{M_{H_3}}{M_{H_2}}=2$, and assume the charged Higgs mass as
$M_{H^+}=600$GeV.}}
\end{center}
\end{figure}
\section{Summary and Conclusion}
Now we summarize our results. 
For the form factors calculation,
we apply the chiral Lagrangian including the vector resonance for the
computation of the scalar and vector form factors of $\tau \to K \pi(\eta) \nu$ decays. 
\begin{itemize}
\item{
We present new counterterms related to the vector mesons
and $\eta_0$ in one-loop level of pseudoscalar mesons 
and show how one can perform the renormalization
in a systematic way for the diagram with arbitrary number of the loop.}
\item{By using the propagator with the one loop corrected self-energy 
of the vector mesons, one can reproduce the vector meson intermediate states.}
\item{We fit our theoretical curve of the hadronic invariant mass distribution 
with that obtained by Belle. By tuning the parameters, we demonstrate that 
one can fit the
hadronic mass distribution up to $ \sim 1300$ MeV well. Between $1300$ MeV and $1500$ MeV
, our prediction is slightly lower than the experimental data.  We also compute the branching fraction $\tau \to K \eta \nu$, which is consistent with the experimental one.}
\end{itemize}
About the CP violation of the two Higgs doublet model,
we study the CP violation of the Higgs sector. The model
was invented to explain the large isospin breaking of bottom and top
due to large $\tan \beta \simeq 40$ \cite{Hashimoto:2004kz}. 
 CP violation of Higgs potential leads to the mixings of CP even and CP odd 
Higgs. The Yukawa couplings of down type quarks and charged leptons with
the neutral Higgs of the second Higgs doublet are large and are CP violating.
We found that;
\begin{itemize}
\item{The CP violation in the neutral Higgs sector leads to the CP violating effect on the quarks and
leptons mass matrices through one loop corrections.}
\item{After removing the CP violating phases in the mass matrices,
one obtains CP violating 
phases of the charged Higgs couplings to the down type quarks and charged leptons.}
\item{The effect is studied in the forward-backward CP asymmetry of $\tau 
\to K \pi \nu$ decay. The order of the asymmetry is $10^{-6} \sim 10^{-7}$. The smallness of the asymmetry comes from the fact that the 
CP violation is loop induced effect.}
\end{itemize}
\begin{acknowledgements}
We would like to thank Dr. D. Epifanov for providing us with the data of Belle.
We would like to thank Dr. H. Takata for providing us with the mathematica
program of the two Higgs doublet model. We also thank K. Nakagawa,
prof. H. Hayashii, Dr. M. Bischolfberger and the members of the physics study 
group of the B factory for fruitful discussion.  
K.Y.L. was supported by the Basic Science Research Program through the NRF
funded by MEST
(2010-0010916).
The work of T. M. is supported by KAKENHI, 
Grant-in-Aid for Scientific Research(C) No.22540283
from JSPS, Japan.
\end{acknowledgements}
\appendix
\section{Derivation of the counterterms using background field method} 
We give the outline of the derivation of the counterterms.
To derive the counterterms, we use the background field method so that
the calculation of the counterterms is consistent with chiral
symmetry \cite{Gasser:1984gg}, \cite{Gasser:1984ux}.
For the purpose, we first write the 
chiral Lagrangian 
in terms of the fields which are decomposed into  
the background fields and the quantum
fields based on Eq.(\ref{eq:lag}).
We decompose the fields into the background field and quantum fluctuation
as follows;
\bea
&&\xi_L=\xi \exp(i \frac{\Delta}{f}),\,
\xi_R=\xi^\dagger \exp(-i \frac{\Delta}{f}),\nn \\
&&\bar{\eta}_0=\eta_0+\Delta^0.
\eea 
where $\xi=\exp(i \frac{\pi}{f})$ and $\pi$ denotes the background pseudoscalar octet fields. $\Delta$ denotes the quantum fluctuation. $\eta_0$ is the background field for singlet pseudoscalar and $\Delta^0$ is its quantum part.  We also introduce
$\bar{\alpha}_\perp$ and $\bar{\alpha}$ defined as
\bea
\bar{\alpha}_{\perp \mu}&=&
\frac{\xi_L^\dagger D_{L\mu} \xi_L -\xi_R^\dagger \partial_\mu \xi_R}{2i}, 
\nn \\
\bar{\alpha}_\mu&=&\frac{\xi_L^\dagger D_{L \mu} \xi_L+ \xi_R^\dagger 
\partial_\mu \xi_R}{2i}.
\eea
Using the notations given above, we write the Lagrangian including
the background field and the quantum parts,
\bea
{\cal \bar{L}}&=&f^2 {\rm Tr} (\bar{\alpha}_{\perp \mu} \bar{\alpha}_{\perp}
^\mu)
+B {\rm Tr}[\xi_R^\dagger M \xi_L+\xi_L^\dagger M \xi_R)] -ig_{2p}{\rm Tr}
(\xi_R^\dagger M \xi_L -\xi_L^\dagger M \xi_R)\cdot \bar{\eta}_0 
\nn \\
&+&\frac{1}{2} \partial_\mu \bar{\eta}_0 \partial^\mu \bar{\eta}_0-\frac{M_0^2}{2}
\bar{\eta}_0^2 + M_{V}^2 {\rm Tr}\left[(V_{\mu}-\frac{\bar{\alpha}_{\mu}}{g})^2 \right]. 
\label{eq:lag2} 
\eea
If we suppress the quantum fluctuation as $\Delta \rightarrow 0$ and $\Delta^0  \rightarrow 0$, then Eq.(\ref{eq:lag2}) equals to Eq.(\ref{eq:lag}).
We note that  
under the chiral transformation, $\xi$ transforms non-linearly as, 
\bea
\xi^\prime&=&g_L \xi h^\dagger= h\xi g_R^\dagger,
\label{eq:tr1}
\eea
while 
$\Delta$ transforms linearly as,
\bea
\Delta^\prime&=& h \Delta h^\dagger.
\label{eq:tr2}
\eea
We also note $\bar{\alpha}$ and $\bar{\alpha}_\perp$ transform as,
\bea
\bar{\alpha}^\prime_\mu&=&h \bar{\alpha}_\mu h^\dagger+ \frac{1}{i} h \partial_\mu h^\dagger, \nn \\
\bar{\alpha}^\prime_{\perp \mu}&=&h \bar{\alpha}_{\perp \mu} 
h^\dagger.
\label{eq:tr3}
\eea
We treat the vector meson $V$ as the background field and it transforms as
\bea
V_\mu^\prime=h V_\mu  h^\dagger + \frac{1}{g i} h \partial_\mu h^\dagger. 
\label{eq:tr4}
\eea
Given the transformations in Eq.(\ref{eq:tr1})-Eq.(\ref{eq:tr4}), 
the Lagrangian of 
Eq.(\ref{eq:lag2}) with 
the chiral breaking term $M$ replaced with the spurion fields 
\cite{Gasser:1984gg} is 
invariant under chiral $SU(3)_L \times SU(3)_R$ transformation.
We next compute the one-loop corrections and identify the divergence.
For the purpose,
one expands ${\cal \bar{L}}$ up to the second order of $\Delta$ and 
$\Delta^0$.
\bea
{\cal \bar{L}}&=&{\cal L}
+ \aTr D_\mu \Delta D^\mu \Delta
+ \left(1-\frac{M_V^2}{g^2 f^2} \right)
 \aTr[\Delta, \alpha_{\perp \mu}][\Delta, {\alpha_\perp}^\mu]\nn \\
&+&i \frac{M_V^2}{g f^2} \aTr(V_\mu-\frac{\alpha_\mu}{g})[\Delta,D^\mu \Delta]
\nn \\
&-& \frac{2B}{f^2}
{\rm \aTr}(\chi_+ \Delta^2)+ \frac{2g_{2p}}{f^2}
{\rm Tr}(\chi_- \Delta^2) \eta_0
+2 \frac{g_{2p}}{f} {\rm \aTr}(\chi_+ \Delta) \Delta^0 
\nn \\
&+&\frac{1}{2} 
\partial_\mu \Delta^0 \partial^\mu \Delta^0 -\frac{M_0^2}{2} \Delta^{02},
\label{eq:soft}
\eea
where $\chi_\pm$ is defined as,
\bea
\chi_+&=&\xi M \xi + \xi^\dagger M \xi^\dagger, \nn \\
\chi_-&=&i(\xi M \xi-\xi^\dagger M \xi^\dagger).
\eea
Since the background fields $(\xi, \eta_0)$ satisfy the
equations of motion
\bea
&& D^{\mu} \alpha_{\perp \mu}+
i \frac{M_V^2}{g f^2}[V^\mu-\frac{\alpha^\mu}{g},\alpha_{\perp \mu}]
=\frac{1}{f^2}(B \chi_-+ g_{2p} \chi_+ \eta_0),
\nn \\ 
&&(\Box +M_{0}^2) \eta_0+ g_{2p} {\rm Tr}(\chi_-)=0,
\label{eq:bfeq}
\eea
the first variation with respect to $\Delta$ and $\Delta^0$ vanishes.
Introducing the octet component field $\Delta^a$ as 
$\Delta=\sum_{a=1}^8 \Delta^a T^a$, 
one can write the quadratic part of
the action in terms of the quantum parts $\Delta^A=(\Delta^0, \Delta^a)$
as,
\bea
\bar{S}=S-\frac{1}{2} \int d^4 x \Delta^A D^{AB} \Delta^B, 
\eea
$D^{AB}$ is a diffrential operator and a $9 \times 9$ matrix for
the nonet space.
\bea
D^{AB}=\begin{pmatrix} (\Box +M_0^2) &  M^{0b}\\
                  M^{a0}      &  D^{ab} \end{pmatrix}, \nn \\
\Gamma^{AB}=\begin{pmatrix} 0 & 0 \\
                            0 & \Gamma^{ab} \end{pmatrix},
v^{AB}=\begin{pmatrix} 0 & 0 \\
                            0 & v^{ab} \end{pmatrix},
\eea
with,
\bea
D^{ab} =(\tilde{d}_\mu \tilde{d}^\mu)^{ab} + \sigma^{ab}-(v^2)^{ab},
\eea
where, 
\bea
&& (\tilde{d}_\mu \Delta)^a=(d_\mu \Delta)^a+v_\mu^{ab} \Delta^b, \nn \\
&& (d_\mu \Delta)^a=\partial_\mu \Delta^a +\Gamma_\mu^{ab} \Delta^b,
\eea
and,
\bea
\Gamma_\mu^{ab}&=&-2i \aTr[T^a,T^b] \alpha_\mu, \nn \\
V_\mu^{ab}&=&-2i \aTr[T^a,T^b] V_\mu,  \nn \\
v^{ab}_\mu&=&\frac{M_V^2}{2 g f^2}
( V^{\mu ab}-\frac{1}{g} \Gamma^{\mu ab}).
\eea
We also define,
\bea
\sigma^{ab}&=&\frac{4B}{f^2} {\rm \aTr}(\chi_+ T^a T^b)
-\frac{4 g_{2p}}{f^2} 
{\rm \aTr}(\chi_- T^a T^b) \eta_0 \nn \\
&-&{ 2 \left(1-\frac{M_V^2}{g^2 f^2}\right) 
{\rm \aTr}[T^a, \alpha_{\perp \mu}][T^b, \alpha_\perp^{\mu}]},
\eea
and
\bea
M^{a0}=M^{0a}=-\frac{2 g_{2p}}{f} 
{\rm Tr}(\chi_+ T^a).
\eea
The effective action including one loop corrections is given by
\bea
&&S_{eff}=S+\Delta S, \nn \\
&& \Delta S=\frac{i}{2} {\rm Tr}{\rm Ln} D^{AB}.
\eea
By introducing, $ 9 \times 9$ matrix, 
\bea
\sigma^{AB}=\begin{pmatrix} M_0^2 & M^{0b} \\
                             M^{a0} & \sigma^{ab} \end{pmatrix},
\eea
one can write $D^{AB}$ as;
\bea
D^{AB} =(\tilde{d}_\mu \tilde{d}^\mu)^{AB} + \sigma^{AB}-(v^2)^{AB}.
\eea
The divergent part of one loop correction can be easily computed with the
heat kernel method \cite{Gasser:1984gg} \cite{Donoghue:1992dd}.
The counterterms can be also obtained with, 
\bea
S_c=-\Delta S|_{div}=\lambda \int 
d^4 x
{\rm Tr}[a_2(x)], \nn \\
\eea
where $a_2(x)$ is given by,
\bea
a_2(x)=\frac{1}{2} \tilde{\sigma}^2+\frac{1}{12}[\tilde{d}_\mu, \tilde{d}^\nu]
[\tilde{d}^\mu,\tilde{d}^\nu]+\frac{1}{6}[\tilde{d}_\mu,[\tilde{d}^\mu,\tilde{\sigma}]],
\eea
where $\tilde{\sigma}=\sigma-v^2$. The trace for $9 \times 9$ matrix can be 
converted to trace for $ 3 \times 3$ matrix which leads to the conterterms of
Eq.(\ref{eq:fullcounter}).
\section{1 loop functions}
Here we summarize the one loop functions which appear in Eq.(\ref{eq:ff1}).
\bea
I_P &=& \int\frac{d^d k}{(2\pi)^d i}\frac{1}{k^2-m_P^2} , \nn \\
\chi_\mu^{QP} &=& \int\frac{d^d k}{(2\pi)^d i}
 \frac{(Q-2k)_\mu}{\{(k-Q)^2-m_Q^2\}(k^2-m_P^2)} ,\nn \\
J_\mu^{QP} &=& \int\frac{d^d k}{(2\pi)^d i}
 \frac{(2k-Q)_\mu (2k-Q)_\nu q^\nu}{\{(k-Q)^2-m_Q^2\}(k^2-m_P^2)},
\label{eq:loopf1}
\eea
By carrying out the loop integral, one obtains,
\bea
I_p&=& -2 m_P^2 \lambda-2 f^2 \mu_P, \nn \\
\chi^{QP}&=&-Q_\mu \frac{\Delta_{PQ}}{Q^2} \bar{J}_{PQ}, \nn \\
J_\mu^{QP}&=&(q_\mu-\frac{Q \cdot q}{Q^2} Q_\mu) \left(-4f^2H_{PQ}+\frac{2}{3} \lambda Q^2-2\lambda \Sigma_{PQ}-2(\mu_Q+\mu_P)f^2 \right) \nn \\
&+& \frac{Q \cdot q}{Q^2} Q_\mu \left(\frac{\Delta_{PQ}^2 \bar{J}_{PQ}}{s}-2\lambda \Sigma_{PQ}-2(\mu_Q+\mu_P) f^2 \right),
\eea
$\bar{J}_{PQ}$ is defined in Eq.(\ref{eq:Jbar}) and Eq.(\ref{eq:Jbar2}). 
$H_{PQ}$ is defined in
Eq.(\ref{eq:HPQ}). 
By taking account of $\eta_0$ and $\eta_8$ mixing, one defines 
$I_{\eta_8}$ and $X^{\eta_8 K}_\mu\ (X_\mu=\chi_\mu,J_\mu)$ 
as,
\bea
I_{\eta_8} &=& I_{\eta}\cos^2\theta_{08} + I_{\eta'}\sin^2\theta_{08} ,\nn \\
X^{\eta_8 K}_\mu &=& X^{\eta K}_\mu\cos^2\theta_{08} 
 + X^{\eta' K}_\mu\sin^2\theta_{08} .
\label{eq:loopf2}
\eea

\section{Two point function of vector mesons}
\label{app_2point}
Let us determine the coefficient of the counterterms
$C_1, C_2$ and $Z_V$ from the renormalization 
for self-energy of vector mesons.
The vector mesons couplings with pseudo scalar mesons
in ${\cal L}$ are,
\bea
{\cal L}^{VPP}&=&-\frac{ M_V^2}{g f^2 i} {\rm Tr}
(V_{\mu} [\Delta, \partial^{\mu} \Delta]) \nn \\
&=& \frac{M_V^2 i}{4 g f^2}
\left[
K^{\ast + \mu} \left(\hat{K}^- \dc_{\mu} \hat{\pi}^0+ 
\sqrt{3}
\hat{K}^- \dc_{\mu} \hat{\eta}_8 + \sqrt{2} \hat{\bar{K}}^0 \dc 
\hat{\pi}^- \right)
\right.   \nn \\
&&+ \left. 2 \rho^{+ \mu}
\left(\hat{\pi}^- \dc \hat{\pi}^0 +\frac{1}{\sqrt{2}} \hat{\bar{K}}^0
\dc \hat{K}^- \right) \right].
\eea
The quantum field for pseudoscalar octet $\Delta$ is denoted by $\hat{\pi}$.
One can parameterize the inverse propagators of vector fields as,
\bea
A_{V}g^{\mu \nu} +B_{V}  Q^{\mu} Q^{\nu}.
\label{eq:invp}
\eea
We study the two point functions for $\rho^+$ and $K^{\ast +}$
mesons.
\bea
A_{V}&=&M_V^2-Q^2+\delta A_{V}, \nn \\
B_{V}&=&1+ \delta B_{V},
\eea
where $\delta A_{V}$ and $\delta B_V$ denote the one loop
corrections including the contribution from the counterterms.
For the $\rho$ meson, they are given as,
\bea
\delta B_{\rho}&=&{Z^r}_V(\mu) + \left(\frac{M_V^2}{g f^2}\right)^2
\left(M^r_{\pi}+\frac{1}{2} M^r_{K} \right), \nn \\
\delta A_{\rho}&=& -Q^2 \delta B_{\rho} -\left(\frac{M_V^2}{ g f^2} 
\right)^2 \left(\mu_{\pi}+\frac{1}{2} \mu_K \right) f^2 \nn \\
&&+ C^r_{1}(\mu) m_{\pi}^2 + C^r_{2}(\mu) (2 m_K^2+ m_\pi^2),
\label{del_rho}
\eea
where 
$\mu_P=\frac{m_P^2}{32 \pi^2 f^2}
 \ln \frac{{m_P}^2}{\mu^2} $.
$M^r_{P}$ ($P=\pi, K$)  are the loop functions for $\pi$ mesons
and $K$ mesons defined as,
\bea
M^r_{P}&=&\frac{1}{12}
\left[\left( 1-\frac{4 m_P^2}{Q^2} \right) \bar{J}_P-\frac{1}{16 \pi^2}
\ln \frac{m_P^2}{\mu^2}-\frac{1}{4 8 \pi^2} \right],\nn \\
\bar{J}_P&=& \left\{ \begin{array}{c}
-\frac{1}{16 \pi^2} \sqrt{1-\frac{4 m_P^2}{Q^2}}
\ln \frac{1+\sqrt{1-\frac{4 m_P^2}{Q^2}}}{1-\sqrt{1-\frac{4 m_P^2}{Q^2}}}
+\frac{1}{8 \pi^2}+ i \frac{1}{16 \pi} \sqrt{1-\frac{4 m_P^2}{Q^2}},
 \quad (Q^2 \geq 4m_P^2), \\
\frac{1}{8 \pi^2} \left(1-\sqrt{\frac{4 m_P^2}{Q^2}-1} \arctan
\frac{1}{\sqrt{\frac{4 m_P^2}{Q^2}-1}}\right),
\quad (Q^2 \leq 4 m_P^2).
\end{array} \right.  
\label{eq:M}
\eea
$Z^r_V(\mu), C^r_1(\mu)$ and $C^r_2(\mu)$ are finite parts
of the renormalization constants defined by,
\bea
Z_V&=&Z^r_V(\mu)-\frac{1}{128 \pi^2}
 \left(\frac{M_V^2}{ g f^2}\right)^2
(C_{UV}+1- \ln \mu^2), \nn \\
C_1&=&C^r_1(\mu)-\frac{3}{128 \pi^2}
 \left(\frac{M_V^2}{ g f^2}\right)^2
(C_{UV}+1- \ln \mu^2), \nn \\
C_2&=&C^r_2(\mu)-\frac{1}{128 \pi^2}
 \left(\frac{M_V^2}{ g f^2}\right)^2
(C_{UV}+1- \ln \mu^2),
\label{ren_con}
\eea
with $C_{UV}$ is the divergent part of
the dimensional regularization,
\bea
C_{UV}=\frac{1}{\epsilon}-\gamma+ \ln(4 \pi),
\eea
where $\epsilon=2-\frac{d}{2}$ and $\gamma$ is Euler constant.
The
 self energy corrections to $K^*$ meson are given as,
\bea
\delta B_{K^*}&=&Z^r_V(\mu) +\frac{3}{4}\left(\frac{M_V^2}{ g f^2} 
\right)^2 (M^r_{K \pi} + M^r_{K \eta_8}), \nn \\
\delta A_K^{\ast}&=&-Q^2 \delta B_{K^{\ast}}+ \frac{3}{4}
\left(\frac{M_V^2}{ g f^2} 
\right)^2  \left[ L_{K \pi}+L_{K \eta_8} -\frac{f^2}{2}(
\mu_{\pi}+2 \mu_K + \mu_{\eta_8}) \right] \nn \\
&&+ C^r_1(\mu) m_K^2 + C^r_2(\mu) (2 m_K^2+ m_\pi^2), 
\label{del_ks}
\eea
where $M^{r}_{PQ}$ and $L_{PQ}$ are the same functions
as the ones defined in Ref. \cite{Gasser:1984gg},
\bea
M^r_{PQ}&=&\frac{1}{12 Q^2} \left(Q^2-2 \Sigma_{PQ}\right) \bar{J}_{PQ}
+ \frac{\Delta_{PQ}^2}{3 Q^4}\left[
\bar{J}_{PQ}-Q^2
\frac{1}{32 \pi^2} \left( \frac{\Sigma_{PQ}}{\Delta_{PQ}^2}+
2 \frac{m_P^2 m_Q^2}{\Delta_{PQ}^3} \ln \frac{m_Q^2}{m_P^2} \right)
\right] \nn \\
&&-\frac{k_{PQ}}{6}+\frac{1}{288 \pi^2}, \nn \\ 
L_{PQ}&=& \frac{\Delta_{PQ}^2}{ 4 s} \bar{J}_{PQ},
\label{eq:ML}
\eea
where $k_{PQ}=\frac{(\mu_{P}-\mu_{Q}) f^2}{\Delta_{PQ}}$.
$\bar{J}_{PQ}$ is a one loop scalar function of pseudo scalar 
mesons with masses $m_P$ and $m_Q$,
Above the threshold; $Q^2 \geq (m_P+m_Q)^2 $, it is given by,
\bea
\bar{J}_{PQ}(Q^2)&=&
\frac{1}{32 \pi^2}
\left[ 2 +\frac{\Delta_{PQ}}{Q^2} \ln 
\frac{m_Q^2}{m_P^2}-\frac{\Sigma_{PQ}}{\Delta_{PQ}} 
\ln \frac{m_Q^2}{m_P^2}\right.\nn \\
&&- \left.
\frac{\nu_{PQ}}{Q^2}
\ln
\frac{(Q^2+\nu_{PQ})^2-\Delta_{PQ}^2}{(Q^2-\nu_{PQ})^2-\Delta_{PQ}^2}
\right]+ \frac{i}{16 \pi} \frac{\nu_{PQ}}{Q^2},
\label{eq:Jbar}
\eea
where,
\bea
\nu_{PQ}^2=Q^4-2 Q^2 \Sigma_{PQ}+ \Delta_{PQ}^2,
\label{eq:nuPQ}
\eea
while below the threshold $(m_P-m_Q)^2 \leq Q^2 \leq (m_P+m_Q)^2$,
\bea
\bar{J}_{PQ}(Q^2)&=&
\frac{1}{32 \pi^2}
\left[ 2 +\frac{\Delta_{PQ}}{Q^2} \ln 
\frac{m_Q^2}{m_P^2}-\frac{\Sigma_{PQ}}{\Delta_{PQ}} 
\ln \frac{m_Q^2}{m_P^2} \right. \nn \\
&&- \left. 2 \frac{\sqrt{-\nu_{PQ}^2}}{Q^2}
\left(
\arctan \frac{Q^2-\Delta_{PQ}}{\sqrt{-\nu_{PQ}^2}}
+\arctan\frac{Q^2+\Delta_{PQ}}{\sqrt{-\nu_{PQ}^2}}\right)
\right],
\label{eq:Jbar2}
\eea
with $\Sigma=m_P^2+m_Q^2$ and $\Delta_{PQ}=m_P^2-m_Q^2$.
\section{Self-energy corrections for $\eta-\eta^\prime$ sector}
Here we have expanded the interaction terms which are needed to compute
$z_{88}, \delta M^2_{08}$ and
$\delta M^2_{88}$ in Eq.(\ref{eq:z88}) 
as well as the wavefunction renomalization
constants $z_K$ and $z_\pi$ in Eq.(\ref{eq:zkp}) within one loop approximation.
The relevant part of the Lagrangian for the calculation is 
\bea
{\cal L}&=&\left(1-\frac{M_V^2}{g^2 f^2} \right)
\aTr[\Delta, \alpha_{\perp \mu}][\Delta, {\alpha_\perp}^\mu]\nn \\
&-& \frac{2 B}{f^2}
{\rm \aTr}(\chi_+ \Delta^2)+ \frac{2 g_{2p}}{f^2}
{\rm Tr}(\chi_- \Delta^2) \eta_0
+2 \frac{g_{2p}}{f} {\rm \aTr}(\chi_+ \Delta) \Delta^0 \nn \\
&=&  \frac{2c}{f^2} \aTr(\Delta \partial_\mu \pi \Delta \partial^\mu \pi
-\Delta^2 \partial_\mu\pi \partial^\mu \pi)- \frac{g_{2p}}{B f}\aTr(\{\pi,\chi\} \Delta^2) \eta_0 \nn \\
&-& \frac{g_{2p}}{2 B f} \aTr(\{\pi,\{\pi, \chi\}\} \Delta) \Delta^0- \frac{1}{2f^2} \aTr(\{\pi,\{\pi, \chi\}\} \Delta^2).\nn \\
\eea
In terms of the component fields, the Lagrangian is,
\def\hk0{\hat{K}^0}
\def\hk0b{\hat{\bar{K}}^0}
\def\hkp{\hat{K}^+}
\def\hkm{\hat{K}^-}
\bea
{\cal L}&=& -\frac{3c}{4 f^2}
\partial \eta_8 \partial \eta_8(\hat{K}^0 \hk0b + \hkp \hkm)-\eta_8 \eta_0  
\frac{g_{2p}}{Bf} \times  \nn \\
&&\Bigr{[}\frac{5 m_\pi^2-8 m_K^2}{6 \sqrt{3}} \hat{\eta}_8 \hat{\eta}_8
+\frac{3 m_\pi^2-4m_K^2}{2\sqrt{3}}(\hat{K}^0\hat{\bar{K}}^0+\hat{K}^+
\hat{K}^-)+\frac{m_\pi^2}{2\sqrt{3}}(\hat{\pi}^0\hat{\pi}^0+
2 \hat{\pi}^+\hat{\pi}^-)\Bigl{]} \nn \\
&-&\frac{g_{2p}}{2 B f} \hat{\eta}_0 \hat{\eta}_8
\{\eta_8^2 \frac{5 m_\pi^2-8 m_K^2}{3 \sqrt{3}} \} \nn \\
&+& \frac{1}{2 f^2} \eta_8^2 \Bigl{[}
\frac{16 m_K^2-7 m_\pi^2}{18} \hat{\eta}_8^2 
+\frac{8m_K^2-3m_\pi^2}{6}(\hat{K}^0\hat{\bar{K}}^0+\hat{K}^+
\hat{K}^-)+\frac{m_\pi^2}{6}(\hat{\pi}^0\hat{\pi}^0+
2 \hat{\pi}^+\hat{\pi}^-)\Bigr{]}\nn \\
&+&\frac{c}{f^2}
\partial K^+ \partial K^- (-\frac{3}{4} \hat{\eta}_8^2
-\frac{\hat{K}^0 \hk0b+2 \hkp \hkm}{2}-
\frac{\hat{\pi}^{02}+2 
\hat{\pi}^+ \hat{\pi}^-}{4}) \nn \\
&+&\frac{c}{f^2}
\partial \pi^+ \partial \pi^- (- \hat{\pi}^{02}
-\hat{\pi}^+ \hat{\pi}^-
-\frac{\hat{K}^0 \hk0b+ \hkp \hkm}{2}). 
\eea
We denote the quantum field for pseudoscalar octet $\Delta$ as $\hat{\pi}$
and singlet $\Delta_0$ as $\hat{\eta_0}$. $\eta_0, K, \pi $ and $\eta_8$ denote
the background field.
The relevant counterterms can be extracted from Eq.(\ref{eq:fullcounter}) as,
\bea
{\cal L}_c&=&\frac{8L_4}{f^2} (2m_K^2+m_\pi^2)
(\partial K^+ \partial K^- +\partial \pi^+ \partial \pi^-+ 
\frac{\partial \eta_8 \partial \eta_8}{2})
+\frac{8L_5}{f^2}(m_K^2\partial K^+ \partial K^- +
m_\pi^2 \partial \pi^+ \partial \pi^-) \nn \\
&+& \frac{8 L_5}{f^2} \frac{4 m_K^2-m_\pi^2}{6} 
\partial \eta_8 \partial \eta_8-\frac{8 L_6}{f^2}
(2m_K^2+m_\pi^2)\frac{4m_K^2-m_\pi^2}{3}\eta_8^2\nn \\
&-&
\frac{16 L_8}{f^2}
(\frac{m_\pi^4}{6}+\frac{(2m_K^2-m_\pi^2)^2}{3}) \eta_8^2 \nn \\
&+& \frac{4 T_3}{\sqrt{3}}\frac{g_{2p}}{Bf} \Delta_{K \pi} (2m_K^2+m_\pi^2) \eta_0 \eta_8 + \frac{8 T_5}{B f} \frac{g_{2p}}{\sqrt{3}} m_K^2 \Delta_{K \pi} \eta_0 \eta_8. 
\eea 
\section{The form factors for $\tau \to K \eta \nu$ decay and $\tau \to K \eta^\prime \nu$ decay}
In this appendix, we give the equations  
of the form factors for $\tau \to K \eta \nu$ and $\tau \to K \eta' \nu$.
In the following equations, $\delta A$ and $\delta B$ imply
$\delta A_{K^\ast}$ and $\delta B_{K^\ast}$, respectively.
In this appendix, we give the equations  
of the form factors for $\tau \to K \eta \nu$ and $\tau \to K \eta' \nu$.
The vector and scalar form factors are given as the sums of the
contribution of 1 PI diagram and $K^\ast$ resonance contribution.
\bea
F_V^{K \eta}&=&F^{1PI}_{V K \eta}+F^{K^\ast}_{V K \eta}, \nn \\
F_S^{K \eta}&=&F^{1PI}_{S K \eta}+F^{K \ast}_{S K \eta}. 
\eea
The contribution of the 1 PI diagrams for $\tau \to K \eta \nu$ 
form factors is computed as,
\bea
&& \langle K^+ \eta|\bar{u_L}\gamma_\mu s_L|0 \rangle|_{1 PI}=\cos\theta_{08}\frac{\sqrt{3}}{\sqrt{2}} \times \nn \\
&& \left(-\frac{1}{2} (1-\frac{M_V^2}{2g^2f^2})q_\mu (\sqrt{z_K z_{88}}-1)
+\frac{3}{16 f^2}\Biggl{[}(1-\frac{M_V^2}{2g^2f^2})^2 (J_\mu^{\pi K}+J_\mu^{\eta_8 K}) \right. \nn \\
&& \left. -c Q_\mu (I_{\eta_8}-2 I_K + I_\pi)-2 c q_\mu 
(I_{\eta_8}+4 I_K + I_\pi) \right. \nn \\
&+& (1-\frac{M_V^2}{2g^2f^2}) \left(\{c(Q^2-\Sigma_{K \eta})+\frac{5m_K^2-3m_\pi^2}{3}\} \chi_{\mu}^{\pi K}+
\{c(Q^2-\Sigma_{K \eta})-\frac{5m_K^2-3m_\pi^2}{9}\} \chi_{\mu}^{\eta_8 K}
\right) \Biggr{]}\nn \\
&-&\frac{q_\mu}{2f^2}\Bigl{[}m_K^2(-K_4 (\frac{M_V^2}{2g^2f^2})^2+8 L_5)
+(2m_K^2+m_\pi^2)(-K_5 (\frac{M_V^2}{2g^2f^2})^2+8 L_4) +\frac{4 L_5}{3}
\Delta_{K \pi} \Bigr{]}\nn \\
&+&\left. L_5 \frac{2}{3 f^2} Q_\mu \Delta_{K \pi}
-\frac{C_5}{4 f^2}(q_\mu Q^2-\Delta_{K \eta} Q_\mu) \right).
\label{eq:etaff}
\eea
Using Eq.(\ref{eq:etaff}), the form factors of 1 PI part is derived as, 
\bea
F^{1PI}_{V K \eta}&=&\sqrt{\frac{3}{2}} \cos \theta_{08}
\Biggl{[}-\left(1-\frac{M_V^2}{2 g^2 f^2}\right)-
\frac{3 c}{2} (H_{K \pi}+H_{K \eta_8})\nn \\
&-&\frac{3}{8}\left(\frac{M_V^2}{g^2f^2}\right)^2
(H_{K \pi}+H_{K \eta_8})-C_5^{r} \frac{Q^2}{2f^2}
+\frac{M_V^2}{2 g^2 f^2} \nn \\
&\times& \Bigl\{-\frac{4}{3}\frac{7 m_K^2-m_\pi^2}{f^2}
L_5^r-\frac{8(2 m_K^2+m_\pi^2)}{f^2} L_4^r+\frac{3 c}{4} (\mu_{\eta_8}+\mu_\pi+
6 \mu_K)\Bigr\}\nn \\
&+& \left(\frac{M_V^2}{2 g^2 f^2}\right)^2
\Bigl\{\frac{m_K^2}{f^2} K_4^r+\frac{2 m_K^2+m_\pi^2}{f^2} K_5^r-\frac{3}{4}(\mu_{\eta_8}+\mu_\pi+2 \mu_K)\Bigr\} \Biggr{]}, \nn \\
F^{1PI}_{S K \eta}&=&
\sqrt{\frac{3}{2}} \frac{\cos \theta_{08}}{Q^2}
\Biggl{[}\left(1-\frac{M_V^2}{2 g^2 f^2}\right)
\Biggl\{
-\Delta_{K \eta}+\frac{3}{8}\Bigl\{
\left(c(Q^2-\Sigma_{K \eta})+ \frac{5m_K^2-3m_\pi^2}{3}\right) 
\frac{\Delta_{K \pi}}{f^2}\bar{J}_{K \pi}\nn \\
&&+\left(c(Q^2-\Sigma_{K \eta})-\frac{5 m_K^2-3 m_\pi^2}{9} \right)
\left(\frac{\Delta_{K \eta}}{f^2}\bar{J}_{K \eta} \cos^2 \theta_{08}
+\frac{\Delta_{K \eta^\prime}}{f^2}\bar{J}_{K \eta^\prime} \sin^2 \theta_{08}
\right) 
\Bigr\} \Biggr\} \nn \\
&+&\frac{3}{8}
\left(1-\frac{M_V^2}{2g^2f^2} \right)^2 \frac{\Delta_{K \eta}}{f^2}
\left(\frac{\Delta_{K \pi}^2}{s}\bar{J}_{K \pi} +
\frac{\Delta_{K \eta}^2}{s}\bar{J}_{K \eta} \cos^2 \theta_{08}
+\frac{\Delta_{K \eta^\prime}^2}{s}\bar{J}_{K \eta^\prime} \sin^2 \theta_{08}
\right) \nn \\
&+&\frac{3c}{4}(\mu_{\eta_8}+\mu_\pi-2\mu_K) Q^2
\Biggr{]}+ 2 \sqrt{\frac{2}{3}} L_5^{r} \frac{\Delta_{K \pi}}{f^2}
\cos \theta_{08} \nn \\
&+& \sqrt{\frac{3}{2}} \cos \theta_{08} \frac{\Delta_{K \eta}}{Q^2}
\frac{M_V^2}{2 g^2 f^2}\Biggl{[} 
\Bigl\{-\frac{4}{3}\frac{7 m_K^2-m_\pi^2}{f^2}
L_5^r-\frac{8(2 m_K^2+m_\pi^2)}{f^2} L_4^r+\frac{3 c}{4} (\mu_{\eta_8}+\mu_\pi+
6\mu_K)\Bigr\}\nn \\
&+& \left(\frac{M_V^2}{2 g^2 f^2}\right)
\Bigl\{\frac{m_K^2}{f^2} K_4^r+\frac{2 m_K^2+m_\pi^2}{f^2} K_5^r-\frac{3}{4}(\mu_{\eta_8}+\mu_\pi+2 \mu_K)\Bigr\}
\Biggr{]}.
\eea
The decay amplitude of the process 
$K^\ast \to K \eta$ is given as,
\bea
T_\mu(K^{\ast+} \to K^+ \eta)=E_{K \eta} q_\mu + Q_\mu \Delta_{K \eta}
{\cal F}_{K \eta}, 
\eea
with
\bea
E_{K \eta}&=& \sqrt{3} \cos \theta_{08} \Biggl{[}
\frac{M_V^2}{4 g f^2}-
\frac{g}{2 M_V^2}(1-\frac{M_V^2}{2 g^2 f^2})
(\delta A + Q^2 \delta B) \nn \\
&& +\frac{M_V^2}{16gf^2}\Bigl\{-3 (\mu_\pi+\mu_{\eta_8}+2 \mu_K)
+c(\mu_\pi+\mu_{\eta_8}+6\mu_K) \nn \\
&-&32 L_4^r \frac{2 m_K^2+m_\pi^2}{f^2}-16 L_5^r
\frac{7m_K^2-m_\pi^2}{3f^2} \Bigr\} \nn \\
&+& \frac{-g}{2 M_V^2}(C_1^r m_K^2+ C_2^r (2 m_K^2+m_\pi^2))
 \nn \\
&-& \frac{g}{2M_V^2}(1-\frac{M_V^2}{2g^2f^2}) (-Q^2)
(\delta B -Z^r_V) +
\frac{C_3^r}{8f^2} Q^2 \Biggr{]} ,\nn \\
{\cal F}_{K \eta} 
&=&\sqrt{3}\cos \theta_{08} \Biggl{[}-\frac{g}{2 M_V^2}(1-\frac{M_V^2}{2g^2f^2})(\delta B-Z^r_V)
-\frac{C_3^r}{8f^2}
\nn \\
&-&\frac{ M_V^2 }{32gf^4}\frac{3c(Q^2-\Sigma_{K \eta})}{Q^2}
\left(\frac{\Delta_{K \pi}}{\Delta_{K \eta}} \bar{J}_{K \pi}
+\bar{J}_{K \eta} \cos^2 \theta_{08} + 
\frac{\Delta_{K \eta^\prime}}{\Delta_{K \eta}} \bar{J}_{K \eta^\prime} 
\sin^2 \theta_{08} \right) 
\nn \\
&-&\frac{ M_V^2 }{32 g f^4}
\frac{5m_K^2-3m_\pi^2}{Q^2} 
\left(\frac{\Delta_{K \pi}}{\Delta_{K \eta}} \bar{J}_{K \pi}
-\frac{1}{3}(
\bar{J}_{K \eta} \cos^2 \theta_{08} + 
\frac{\Delta_{K \eta^\prime}}{\Delta_{K \eta}} \bar{J}_{K \eta^\prime} 
\sin^2 \theta_{08})\right) 
\Biggr{]}.
\eea
Using the $K^\ast \to K \eta$ decay amplitude, the contribution to
the form factor is given by,
\bea
F_{V K \eta}^{K^\ast}&=&-2 E_{K \eta} \frac{G+ Q^2 \cal{H}}{M_V^2+\delta A}
,\nn \\
F_{S K \eta}^{K^\ast}&=&-2 G \frac{\Delta_{K \eta}}{Q^2} 
\frac{E_{K \eta}+Q^2 {\cal F}_{K \eta}}{M_V^2+\delta A+Q^2 \delta B}.
\eea
The form factor for $\tau \to K \eta^\prime \nu$ is also given as,
\bea
F_V^{K \eta^\prime}=F^{1PI}_{V K \eta^\prime}+F^{K^\ast}_{V K \eta^\prime},
\nn \\
F_S^{K \eta^\prime}=F^{1PI}_{S K \eta^\prime}+F^{K^\ast}_{S K \eta^\prime}.
\eea
\bea
F^{1PI}_{V K \eta^\prime}&=&\sqrt{\frac{3}{2}} \sin \theta_{08}
\Biggl{[}-\left(1-\frac{M_V^2}{2 g^2 f^2}\right)-
\frac{3 c}{2} (H_{K \pi}+H_{K \eta_8})\nn \\
&-&\frac{3}{8}\left(\frac{M_V^2}{g^2f^2}\right)^2
(H_{K \pi}+H_{K \eta_8})-C_5^{r} \frac{Q^2}{2f^2}
+\frac{M_V^2}{2 g^2 f^2} \nn \\
&\times& \Bigl\{-\frac{4}{3}\frac{7 m_K^2-m_\pi^2}{f^2}
L_5^r-\frac{8(2 m_K^2+m_\pi^2)}{f^2} L_4^r+\frac{3 c}{4} (\mu_{\eta_8}+\mu_\pi+
6 \mu_K)\Bigr\}\nn \\
&+& \left(\frac{M_V^2}{2 g^2 f^2}\right)^2
\Bigl\{\frac{m_K^2}{f^2} K_4^r+\frac{2 m_K^2+m_\pi^2}{f^2} K_5^r-\frac{3}{4}(\mu_{\eta_8}+\mu_\pi+2 \mu_K)\Bigr\} \Biggr{]} \nn \\
F^{1PI}_{S K \eta^\prime}&=&
\sqrt{\frac{3}{2}} \frac{\sin \theta_{08}}{Q^2}
\Biggl{[}\left(1-\frac{M_V^2}{2 g^2 f^2}\right)
\Biggl\{
-\Delta_{K \eta^\prime}+\frac{3}{8}\Bigl\{
\left(c(Q^2-\Sigma_{K \eta^\prime})+ \frac{5m_K^2-3m_\pi^2}{3}\right) 
\frac{\Delta_{K \pi}}{f^2}\bar{J}_{K \pi}\nn \\
&&+\left(c(Q^2-\Sigma_{K \eta^\prime})-\frac{5 m_K^2-3 m_\pi^2}{9} \right)
\left(\frac{\Delta_{K \eta}}{f^2}\bar{J}_{K \eta} \cos^2 \theta_{08}
+\frac{\Delta_{K \eta^\prime}}{f^2}\bar{J}_{K \eta^\prime} \sin^2 \theta_{08}
\right) 
\Bigr\} \Biggr\} \nn \\
&+&\frac{3}{8}
\left(1-\frac{M_V^2}{2g^2f^2} \right)^2 \frac{\Delta_{K \eta^\prime}}{f^2}
\left(\frac{\Delta_{K \pi}^2}{s}\bar{J}_{K \pi} +
\frac{\Delta_{K \eta}^2}{s}\bar{J}_{K \eta} \cos^2 \theta_{08}
+\frac{\Delta_{K \eta^\prime}^2}{s}\bar{J}_{K \eta^\prime} \sin^2 \theta_{08}
\right) \nn \\
&+&\frac{3c}{4}(\mu_{\eta_8}+\mu_\pi-2\mu_K) Q^2
\Biggr{]}+ 2 \sqrt{\frac{2}{3}} L_5^{r} \frac{\Delta_{K \pi}}{f^2}
\sin \theta_{08} \nn \\
&+& \sqrt{\frac{3}{2}} \sin \theta_{08} \frac{\Delta_{K \eta^\prime}}{Q^2}
\frac{M_V^2}{2 g^2 f^2}\Biggl{[} 
\Bigl\{-\frac{4}{3}\frac{7 m_K^2-m_\pi^2}{f^2}
L_5^r-\frac{8(2 m_K^2+m_\pi^2)}{f^2} L_4^r+\frac{3 c}{4} (\mu_{\eta_8}+\mu_\pi+
6\mu_K)\Bigr\}\nn \\
&+& \left(\frac{M_V^2}{2 g^2 f^2}\right)
\Bigl\{\frac{m_K^2}{f^2} K_4^r+\frac{2 m_K^2+m_\pi^2}{f^2} K_5^r-\frac{3}{4}(\mu_{\eta_8}+\mu_\pi+2 \mu_K)\Bigr\}
\Biggr{]}.
\eea
The decay amplitude of the process 
$K^\ast \to K \eta^\prime$ is given as,
\bea
T_\mu(K^{\ast+} \to K^+ \eta^\prime)=E_{K \eta^\prime} 
q_\mu + Q_\mu \Delta_{K \eta^\prime}
{\cal F}_{K \eta^\prime},
\eea
with
\bea
E_{K \eta^\prime}&=& \sqrt{3} \sin \theta_{08} \Biggl{[}
\frac{M_V^2}{4 g f^2}-
\frac{g}{2 M_V^2}(1-\frac{M_V^2}{2 g^2 f^2})
(\delta A + Q^2 \delta B) \nn \\
&& +\frac{M_V^2}{16gf^2}\Bigl\{-3 (\mu_\pi+\mu_{\eta_8}+2 \mu_K)
+c(\mu_\pi+\mu_{\eta_8}+6\mu_K) \nn \\
&-&32 L_4^r \frac{2 m_K^2+m_\pi^2}{f^2}-16 L_5^r
\frac{7m_K^2-m_\pi^2}{3f^2} \Bigr\} \nn \\
&+& \frac{-g}{2 M_V^2}(C_1^r m_K^2+ C_2^r (2 m_K^2+m_\pi^2))
 \nn \\
&-& \frac{g}{2M_V^2}(1-\frac{M_V^2}{2g^2f^2}) (-Q^2)
(\delta B -Z^r_V) +
\frac{C_3^r}{8f^2} Q^2 \Biggr{]} ,\nn \\
{\cal F}_{K \eta^\prime} 
&=&\sqrt{3}
\sin \theta_{08} \Biggl{[}-\frac{g}{2 M_V^2}(1-\frac{M_V^2}{2g^2f^2})(\delta B-Z^r_V)
-\frac{C_3^r}{8f^2}
\nn \\
&-&\frac{M_V^2 }{32gf^4}\frac{3c(Q^2-\Sigma_{K \eta^\prime})}{Q^2}
\left(\frac{\Delta_{K \pi}}{\Delta_{K \eta^\prime}} \bar{J}_{K \pi}
+
\frac{\Delta^{K \eta}
\bar{J}_{K \eta}}{\Delta_{K \eta^\prime}} 
\cos^2 \theta_{08} + 
\bar{J}_{K \eta^\prime} 
\sin^2 \theta_{08} \right) 
\nn \\
&-&\frac{M_V^2 }{32 g f^4}
\frac{5m_K^2-3m_\pi^2}{Q^2} 
\left(\frac{\Delta_{K \pi}}{\Delta_{K \eta^\prime}} \bar{J}_{K \pi}
-\frac{1}{3}(
\frac{\Delta^{K \eta}
\bar{J}_{K \eta}}{\Delta^{K \eta^\prime}} \cos^2 \theta_{08} + 
\bar{J}_{K \eta^\prime} 
\sin^2 \theta_{08})\right) 
\Biggr{]}.
\eea
Using the $K^\ast \to K \eta^\prime$ decay amplitude, the contribution to
the form factor is given by,
\bea
F_{V K \eta^\prime}^{K^\ast}&=&
-2 E_{K \eta^\prime} \frac{G+ Q^2 \cal{H}}{M_V^2+\delta A}
,\nn \\
F_{S K \eta}^{K^\ast}&=&-2 G \frac{\Delta_{K \eta^\prime}}{Q^2} 
\frac{E_{K \eta^\prime}+Q^2 {\cal F}_{K \eta^\prime}}{M_V^2+\delta A+Q^2 \delta B}.
\eea

\end{document}